\documentclass[iop,revtex4-1]{emulateapj}
\usepackage{subfigure,graphicx}
\usepackage{url}
\usepackage[bookmarks=false]{hyperref}
\usepackage{longtable}
\usepackage{appendix}
\usepackage[T1]{fontenc}
\usepackage{graphicx}
\usepackage[outdir=./]{epstopdf}
\usepackage{graphics}
\usepackage{placeins}
\usepackage{float}
\usepackage{xspace}
\usepackage{afterpage}
\hypersetup{
  colorlinks= true, 
  urlcolor  = black, 
  linkcolor = red, 
  citecolor = blue 
}

\newcommand{\herschel}{\textit {Herschel}\xspace}

\newcommand{\ci}{[\ion{C}{1}]\xspace}
\newcommand{\cii}{[\ion{C}{2}]\xspace}

\newcommand{\oi}{[\ion{O}{1}]\xspace}
\newcommand{\oiii}{[\ion{O}{3}]\xspace}
\newcommand{\nii}{[\ion{N}{2}]\xspace}

\newcommand{\hii}{\ion{H}{2}\xspace}

\shorttitle{Neutral and Molecular Gas in Early Type Galaxies}   
\shortauthors{Lapham and Young} 

\begin{document}

\title{SPIRE Spectroscopy of Early Type Galaxies}

\author{Ryen Carl Lapham and Lisa M. Young}
\affil{Physics Department, New Mexico Institute of Mining and Technology, 801 Leroy Place, Socorro, NM 87801; ryen.lapham@student.nmt.edu, lyoung@physics.nmt.edu}


\begin{abstract}
We present SPIRE spectroscopy for 9 early-type galaxies (ETGs) representing the most CO-rich and far-infrared (FIR) bright galaxies of the volume-limited Atlas3D sample. Our data include detections of mid to high J CO transitions (J=4-3 to J=13-12) and the \ci (1-0) and (2-1) emission lines. CO spectral line energy distributions (SLEDs) for our ETGs indicate low gas excitation, barring NGC 1266. We use the \ci emission lines to determine the excitation temperature of the neutral gas, as well as estimate the mass of molecular hydrogen. The masses agree well with masses derived from CO, making this technique very promising for high redshift galaxies. We do not find a trend between the \nii 205 flux and the infrared luminosity, but we do find that the \nii 205/CO(6-5) line ratio is correlated with the 60/100 \micron\ Infrared Astronomical Satellite (IRAS) colors. Thus the \nii 205/CO(6-5) ratio can be used to infer a dust temperature, and hence the \textbf{intensity} of the interstellar radiation field (ISRF). Photodissociation region (PDR) models show that use of \ci and CO lines in addition to the typical \cii, \oi, and FIR fluxes drive the model solutions to higher densities and lower values of G$_0$. In short, the SPIRE lines indicate that the atomic and molecular gas in the CO-rich ETGs have similar properties to other galaxies. As might be expected from their low levels of star formation activity, the ETGs have rather low excitation CO SLEDs, low temperatures inferred from the \ci lines, and modestly lower \ci/CO ratios. 

\end{abstract}


\section{Introduction}
\label{sec:intro}

Molecular gas is an important driver of galaxy evolution.  Star formation is dependent upon reservoirs of cool molecular gas, and in turn controls nearly every observable property of galaxies, from the infrared (IR) to the  ultraviolet (UV). Though it is molecular hydrogen that fuels star formation, the lack of a permanent electric dipole moment makes it difficult to detect the bulk of the cold H$_2$ in emission, so we traditionally use $^{12}$CO (CO) and its many variants to infer properties of molecular gas such as mass, temperature, and its dynamics.  Ground based observations of the CO(1-0) emission line have been used extensively to estimate the total gas content of galaxies, however, higher transition lines are difficult to access with ground based telescopes because the Earth's atmosphere is opaque at their wavelengths.  With the launch of the \herschel Space Observatory \citep{Pil10} mid-J and high-J lines became accessible for observation in nearby galaxies.  The Spectral and Photometric Imaging Receiver (SPIRE) instrument aboard \herschel is capable of observing the CO ladder from J=4-3 up to J=13-12, and even higher transitions with the Photodetector Array Camera and Spectrometer (PACS) instrument, allowing the study of warmer (but still molecular) gas.  Though this warm molecular gas accounts for only about 10\% of the mass of molecular gas, it accounts for roughly 90\% of the total CO luminosity \citep{Kam2014}.  Combining ground based observations of low-J CO transitions with SPIRE observations allows the creation of CO spectral line energy distributions (SLEDs), which provide information on sources of heating other than collisional excitation.

In addition to the bevy of CO transitions in the SPIRE data, there are a pair of neutral carbon emission lines. Neutral carbon was originally thought to exist in a thin photodissociation region (PDR) layer between molecular gas and the dissociation front where hydrogen transitions from molecular to atomic, based on early work with PDR models \citep{Tielens}. However, \citet{Keene} found a peak in \ci abundance deep inside a molecular cloud, contradicting the PDR model's picture of a surface layer of \ci outside molecular layers, leading to the proposal of clumpy molecular clouds.  Since then, many more observations have found evidence that neutral carbon exists alongside CO, providing support for the notion that molecular clouds are comprised of many clumps (i.e. \citet{Ikeda, Krips}). \citet{PapaGreve} present arguments that \ci and CO are concomitant, and measure molecular gas mass using only the \ci (1-0) transition. More recently \citet{Kam2014} used the \ci (1-0) and (2-1) transitions and Boltzmann distributions to estimate the total gas content in nearby galaxies. \ci is particularly useful as a tracer of molecular hydrogen in galaxies with low metallicities and/or increased UV radiation fields.  In such galaxies the dissociation of CO reduces its reliability as a tracer of molecular hydrogen, while neutral carbon remains just as powerful.  This is of particular import to high redshift galaxies, many of which have low metallicities and higher star formation rates. Conveniently, the \ci lines can be more easily observed at their shifted frequencies for high redshift sources. Additionally, the \ci lines can be used to determine the temperature of neutral gas, and can be used in conjunction with CO emission lines to constrain the density of molecular gas.

These observations are some of the first of their kind for early-type galaxies (ETGs), and will provide insight into multiple phases of the interstellar medium (ISM) in ETGs, particularly how they compare to other morphological types. In Section \ref{sec:data} we outline the sample properties, the observations, and the data reduction.  In Section \ref{sec: ladders} we consider the excitation of the molecular gas and possible excitation mechanisms, as well as the relations between each CO transition and star formation.  In Section \ref{sec: carbon} we explore the use of the \ci emission line ratio as a temperature probe and test the possibility of deriving H$_2$ masses with the \ci emission lines.  In Section \ref{sec:budget} we look at how the cooling efficiency of CO and \ci change as a function of dust temperature.  In Section \ref{sec:pdrt} PDR models are used to infer the volume averaged density and interstellar radiation field (ISRF) for ETGs and the results are compared to previous values found for spiral galaxies.  Lastly, in Section \ref{sec:CICOSFR} RADEX models are used to estimate the temperature and density of H$_2$.

\section{Data}
\label{sec:data}

\subsection{Sample Selection}
\label{sec:selection}
The nine galaxies chosen for observation are a subset of the most far infrared (FIR) bright members of the Atlas3D sample that additionally have the largest CO(1-0) fluxes \citep{Cappellari, Young2011}. This choice was made to ensure as many detections as possible for the various CO transitions, and to complement PACS observations of the \cii, \oi, and \nii 122 emission lines discussed in \citep{Lapham}.  

Atlas3D was a volume-limited sample of high mass elliptical and lenticular galaxies within 40 Mpc.  The sample selection was based on K-band magnitute and stellar morphology (lack of spiral arms), but not color, so the sample includes some gas-rich ETGs actively forming stars.  The FIR selection for this project was done using Infrared Astronomical Satellite (IRAS) fluxes.  The resulting subset of galaxies are mostly red sequence galaxies, with stellar masses 2$\cdot$10$^{10}$ to 2$\cdot$10$^{11}$ M$_{\odot}$, though a couple are as blue as typical spirals (see Figure 1 in \citet{Lapham}).

The bluest galaxies in the sample are NGC 1222, 2764, and 7465, with NUV-K and u-r colors similar to spiral galaxies--clear evidence of recent star formation activity \citep{Young2014, McDermid}.  NGC 1222 and NGC 7465 are also recent merger remnants, and they contain kinematically disturbed molecular gas which has not yet settled into relaxed disks \citep{Alatalo}.  The interactions have been driving high levels of star formation activity and high FIR luminosities.  NGC 1266 is also unusual for having an active galactic nucleus (AGN) driven molecular outflow \citep{Alatalo2011}.  The remaining galaxies in the sample evidently have kinematically relaxed molecular disks, to the best of our knowledge.  NGC 3665 has powerful radio jets \citep{Nyland}, but they are apparently not interacting with the molecular disk at a level we can detect.


\subsection{Observation and Reduction}
\label{sec:observations}
All galaxies in the sample were observed by the SPIRE instrument \citep{Griff} aboard the \herschel space-borne observatory \citep{Pil10} with eight observations coming from the project OT1\_lyoung\_1 and the NGC 1266 obsevation coming from OT1\_jglenn\_1.  SPIRE has a three-band imaging photometer, as well as a Fourier Transform Spectrometer (FTS).  Broadband photometric data is taken at roughly 250, 350, and 500 \micron, while the FTS covers 194-671 \micron\ (447-1550 GHz) with two bands (SPIRE short wavelength and long wavelength, SSW and SLW).  The spectral resolution is 1.2 GHz in high resolution mode, and the beam FWHM ranges from 17-42\arcsec. Previous CO observations from \citet{Alatalo} suggested the molecular emission would be mostly or entirely contained within the beam of the central SPIRE detector, so observations were done as single pointings in the high spectral resolution mode.  Additionally, all of the galaxies have SPIRE photometric observations available  to assist with calibrating the spectroscopic observations.

\begin{figure*}
\epsscale{1.0 } 
\plotone{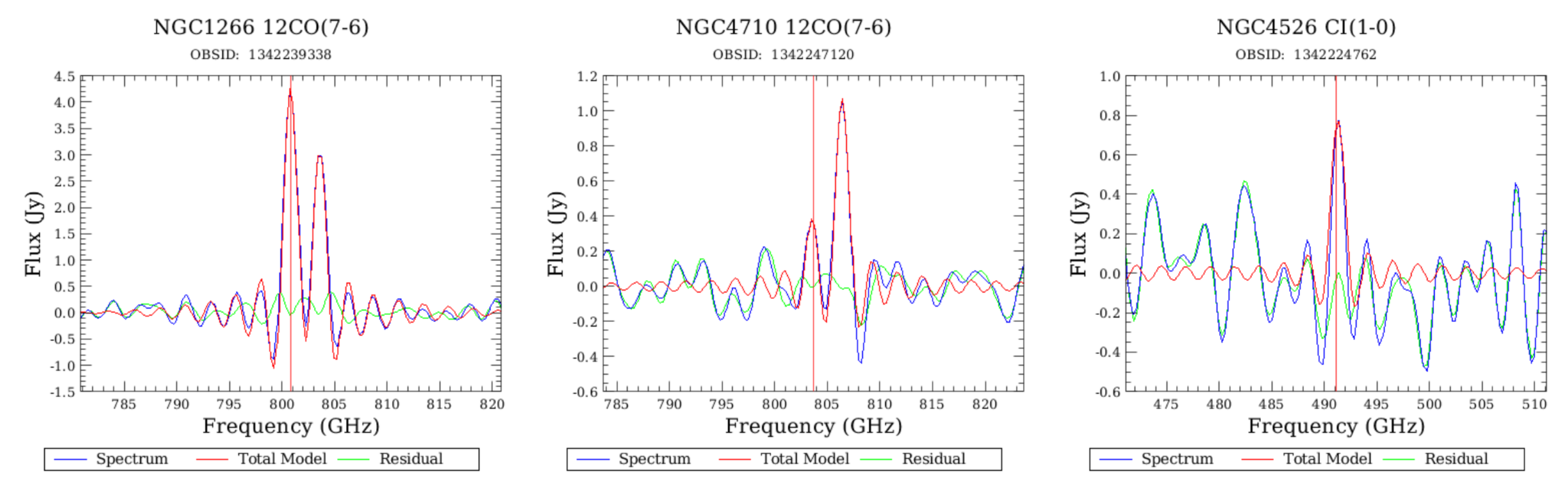} 
\caption{A few example spectra from the SPIRE observations.  The strong line to the right of CO(7-6) line in the first two panels is the \ci(2-1) emission line. A vertical line has been drawn at the expected line center for each of the lines.  The full spectra for each of our nine galaxies can be found in \citet{Lapham}.}
\label{SPIRE_spectra}
\end{figure*}

The SPIRE spectra were reduced in the Herschel Interactive Processing Environment (HIPE) 14, using the built in Single Pointing user pipeline \citep{Ott10}.   In addition to the standard reduction, we also processed the spectra with HIPE's Semi-Extended Flux Correction Tool (SECT).   Because the central detector might not actually cover the entire source, the SECT is used to create a total flux for each source, as well as set a standard reference beam since the beam size is strongly wavelength dependent.  The SECT can optimize the size of the source based on the overlap between the SSW and SLW spectra, however the spectra have random continuum offsets (with respect to each other) on the order of 0.3-0.4 Jy \citep{Hopwood}, and in our sources the offsets are a large percentage of the flux in the overlapping region.  Instead we prefer to specify the source size, position angle, and eccentricity based on two dimensional Gaussian models of the CO (1-0) emission. Additionally, we do not require that our final spectrum match the photometry perfectly, but rather use the photometry as a sanity check, along with prior knowledge of the probable emission size. The spectra before and after the SECT can be seen in Figure 4 of \citet{Lapham}.  A few examples of emission lines from the SPIRE spectra can be seen in Figure \ref{SPIRE_spectra}. 

\subsection{Measuring Line Fluxes}
\label{sec:lineflux}
Once the SPIRE spectra were reduced in the pipeline and corrected by the SECT, the CO and \ci emission lines were fit simultaneously by HIPE's built in Spectrometer Line Fitting script.  We used the unapodized spectra in the fitting routines, and fit a sinc model to each line simultaneously.  Uncertainties were calculated using a Monte Carlo method explained in detail in the appendix.  \textbf{Since the \nii 205 \micron\ lines can be broader than the instrumental profile, those lines were fit with a sinc-Gauss or two sinc components in cases where the CO(1-0) spectra had sharply double-horned shapes.  For NGC 3665 and NGC 4526 in particular (see Figure 8 of \citet{Lapham}) the \nii 205 \micron\ were better fit with two sinc components rather than one sinc-Gauss, however, the differences in the derived flux measurements were marginal.}  The fluxes and uncertainties for the \nii 205 \micron\ line were previously published in \citet{Lapham}.  \href{http://herschel.esac.esa.int/Docs/SPIRE/html/spire_om.html}{The SPIRE Handbook version 2.5} recommends summing the absolute calibration uncertainty (4\%) and relative calibration uncertainty (1.5\%) directly for a conservative total estimate of the calibration uncertainty of 5.5\%.   We have not combined the calibration uncertainties with our Monte Carlo uncertainties, as they are typically only a small component of the error budget.  Fluxes and uncertainties for the \ci and CO emission lines can be seen in Table \ref{COCIfluxes}.  Correction factors applied by the SECT can be seen in Table \ref{CorrFactors}. 

Most of the CO and \ci lines were previously measured by \citet{Kam2016}.  When comparing to their work, we find our CO fluxes to be on average 21\% greater.  The \ci lines are in better agreement, averaging 17\% greater than those in \citet{Kam2016}.  We prefer to use a more direct approach to measure the line fluxes after correcting with the SECT, rather than assuming that the spatial distribution of the gas can be inferred from the FIR continuum levels as \citet{Kam2016} did.  Despite these differences, we find similar results.

\begin{turnpage}
\begin{deluxetable*}{ lcccccccccccc }
\tablecaption{CO and \ci fluxes}
\tablehead{
 \colhead{Galaxy} & \colhead{CO(4-3)} & \colhead{CO(5-4)} & \colhead{CO(6-5)} & \colhead{CO(7-6)} & \colhead{CO(8-7)} & \colhead{CO(9-8)} & \colhead{CO(10-9)} & \colhead{CO(11-10)} & \colhead{CO(12-11)} & \colhead{CO(13-12)} & \colhead{\ci(1-0)} & \colhead{\ci(2-1)} } 
\tablecomments{Fluxes and uncertainties for the CO and \ci transitions.  For lines with fluxes below the 3$\sigma$ level, a 3$\sigma$ upper limit is given.  All values are in units of 10$^{-18}$ W m$^{-2}$.}
\startdata

NGC 1222 &$<$ 11.3&13 $\pm$ 2.1&11.8 $\pm$ 1.3&11.4 $\pm$ 1.0&8.4 $\pm$ 2.4&5.4 $\pm$ 1.5&$<$ 3.8&$<$ 3.8&$<$ 5.5&$<$ 12.2&9.4 $\pm$ 3.1&8.6 $\pm$ 0.7\\
NGC 1266 &29.9 $\pm$ 4.5&42.7 $\pm$ 2.8&43.7 $\pm$ 1.4&48.5 $\pm$ 1.8&50.1 $\pm$ 2.0&42.8 $\pm$ 2.6&35.3 $\pm$ 5.3&31.7 $\pm$ 2.8&30.4 $\pm$ 1.9&20.3 $\pm$ 2.1&15.6 $\pm$ 3.2&32.1 $\pm$ 1.5\\
NGC 2764 &17.7 $\pm$ 3.2&9.1 $\pm$ 2.6&4.7 $\pm$ 1.1&4.4 $\pm$ 1.0&$<$ 5.2&$<$ 5.1&$<$ 4.4&$<$ 5.8&$<$ 8.5&$<$ 10.5&$<$ 4.9&7.1 $\pm$ 0.9\\
NGC 3665 &$<$ 7.0&$<$ 5.7&$<$ 3.5&$<$ 2.8&$<$ 4.3&$<$ 3.7&$<$ 5.0&$<$ 3.1& $<$ 10.0&$<$ 7.8&$<$ 5.8&3.8 $\pm$ 0.8\\
NGC 4459 &$<$ 6.8&$<$ 6.4&3.6 $\pm$ 0.8& $<$ 2.3 & $<$ 3.8 &$<$ 5.3&$<$ 3.5& $<$3.8 &$<$ 3.2&$<$ 5.5&$<$ 8.9&3.1 $\pm$ 0.8\\
NGC 4526 &12.4 $\pm$ 2.9&7.6 $\pm$ 2.1&5.8 $\pm$ 1.3&$<$ 3.8&$<$ 5.4&$<$ 4.6&$<$ 4.0&$<$ 4.3&$<$ 4.6&$<$ 7.2&9.0 $\pm$ 2.5&7.0 $\pm$ 1.0\\
NGC 4710 &15.7 $\pm$ 4.0&17.2 $\pm$ 3.3&7.1 $\pm$ 1.1&$<$ 3.2&$<$ 10.4&$<$ 8.8&6.4 $\pm$ 1.8&$<$ 9.3& $<$ 4.5 & $<$ 9.3 &$<$ 8.6&12.3 $\pm$ 1.1\\
NGC 5866 &12.5 $\pm$ 2.8&6.6 $\pm$ 1.8&7.3 $\pm$ 1.1&4.0 $\pm$ 1.0&$<$ 6.9&$<$ 6.9&$<$ 4.3&$<$ 3.7& $<$ 13.9 &$<$ 15.9& $<$ 16.6 &9.0 $\pm$ 0.8\\
NGC 7465 &$<$ 29.0&$<$ 21.0&$<$ 6.3&$<$ 12.1&$<$ 11.5&$<$ 8.5&$<$ 16.6&$<$ 16.7&$<$ 16.8& $<$ 31.1 &$<$ 38.7&$<$ 10.2

\enddata
\label{COCIfluxes}
\end{deluxetable*}
\end{turnpage}

\section{Analysis}
\label{sec:analysis}


\subsection{CO Ladders}
\label{sec: ladders}
CO has long been recognized as a powerful tracer of otherwise undetectable molecular gas, specifically its ground state J=1-0 transition.  However, with the launch of the Herschel satellite, many mid to high J CO transitions previously obscured by the atmosphere could be observed.  The SPIRE spectral range contains the CO transitions J=4-3 to J=13-12 for nearby galaxies.  These higher transitions allow us to gain insight on a different phase of molecular gas, a phase that is warm (relative to J=1-0 emitting gas) but still molecular. For instance, the CO(1-0) and (2-1) transitions have excitation temperatures of 5.53 K and 16.6 K, respectively, while the mid J transitions (4-3), (5-4), and (6-5) have excitation temperatures of 55.32 K, 82.97 K, and 116.16 K, respectively, and the high J lines (11-10), (12-11), and (13-12) have excitation temperatures of 364.97 K, 431.29 K, and 503.13 K, respectively.  Which transition has the largest flux (where the CO SLED peaks) depends on the excitation mechanism of the gas (i.e. collisions, shocks, X-rays) \citep{Rosenberg}. 

\begin{figure*}
\epsscale{1.0 } 
\plotone{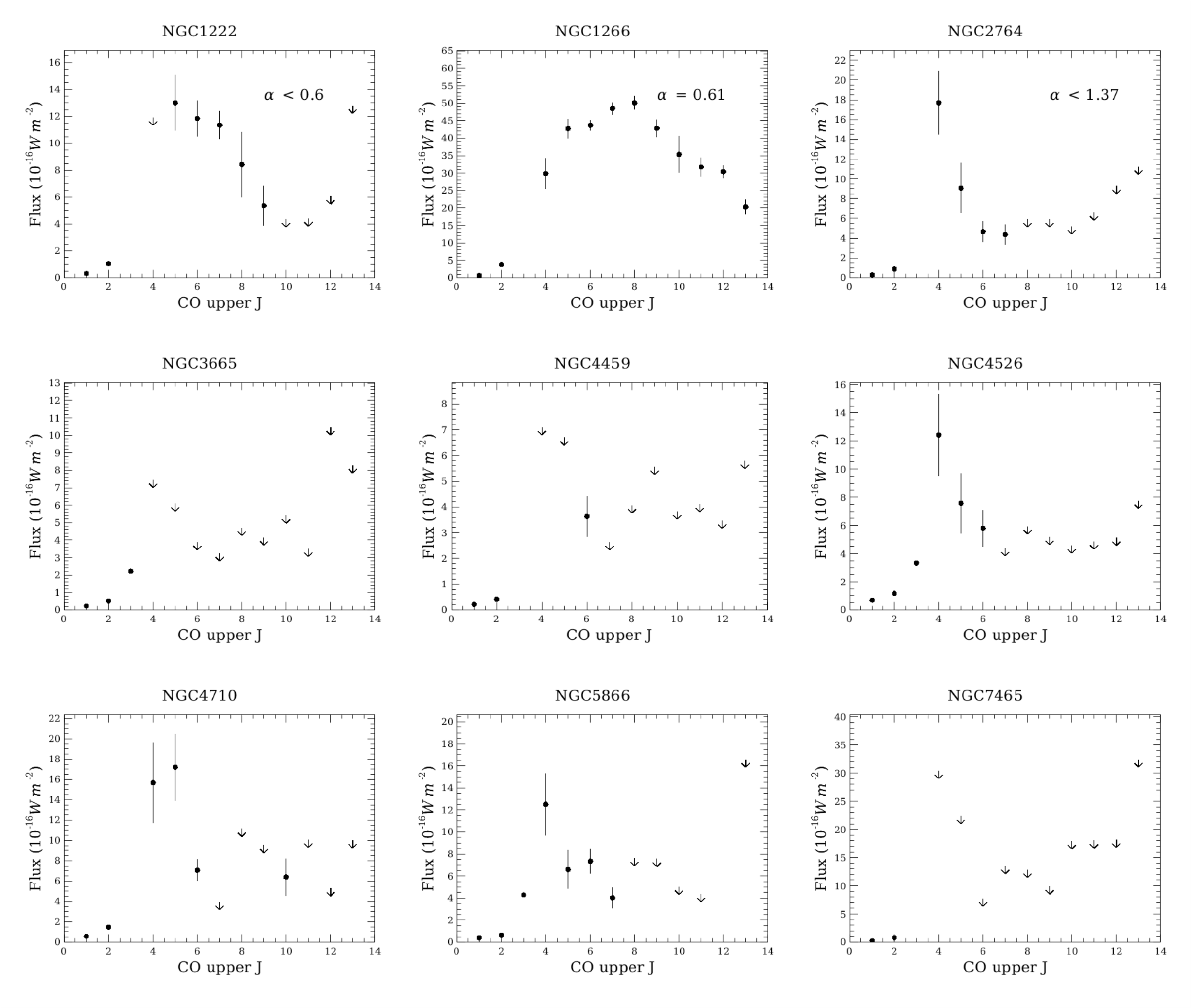} 
\caption{CO ladders for each of our galaxies. The excitation parameter $\alpha$ is shown in the upper right corner for galaxies with enough data to compute the parameter. }
\label{COladders}
\end{figure*}

The CO ladders for each of our galaxies can be seen in Figure \ref{COladders}. Observations of CO(1-0) and CO(2-1) with the Institute for Radio Astronomy in the Millimeter Range (IRAM) 30 m telescope are taken from \citet{Young2011}, with a few coming from \citet{Young2008}, \citet{Crocker2012}, and \citet{Welch}. Observations of CO(3-2) with the James Clerk Maxwell Telescope (JCMT) were available for three of our galaxies (NGC 3665, 4526 and 5866) \citep{Bayet}. 

\subsubsection{Excitation based on Ladder Shape}
\label{sec: shape}
In a study of luminous infrared galaxies (LIRGs) and ultra-luminous infrared galaxies (ULIRGs), \citet{Rosenberg} separate CO ladders into 3 distinct classes by a parameter $\alpha$ that they define as (CO(11-10) + CO(12-11) + CO(13-12))/(CO(5-4) + CO(6-5) + CO(7-6)).  Class I are galaxies with $\alpha$ $<$ 0.33, Class II have 0.33 $<$ $\alpha$ $<$ 0.66, and Class III have $\alpha$ $>$ 0.66.  The higher excitation classes (II and III) require some form of heating other than UV radiation, whether it be mechanical, X-ray, or others. Roughly speaking class II require mechanical, class III require X-rays, but there are other necessary criteria (i.e. line widths) to fully characterize the heating mechanism.  See \citet{Rosenberg} for more details.  

Unfortunately we do not have detections for all of these lines in all of our galaxies, so we can only determine upper limits for some, while others have too few detections to compute an alpha value.  If we had detections for neighboring CO lines a linear interpolation was used to define a flux value.  According to this classification, of our nine galaxies, NGC 1222 and 2764 have upper limits of $\alpha$ $<$ 0.6 and $\alpha$ $<$ 1.37, respectively, while NGC 1266 has a value of $\alpha$ = 0.61.  This places NGC 1266 at the upper end of Class II.   NGC 1266 is an unusual ETG, and has been previously studied in detail by \citet{Glenn, Pellegrini}.

The upper J level at which the CO SLEDs peak is also a crude approximation of the gas excitation, and we find that NGC 2764, 3665, 4526, 5866, and 7465 peak at an upper J of 4, while NGC 1222 and 4710 peak at J=5, and NGC 1266 peaks at J=8.  Upper limits make it difficult to define the peak for NGC 4459, but it appears to peak at an upper J of 4 or 5.  

\begin{figure}
\epsscale{1.0 } 
\plotone{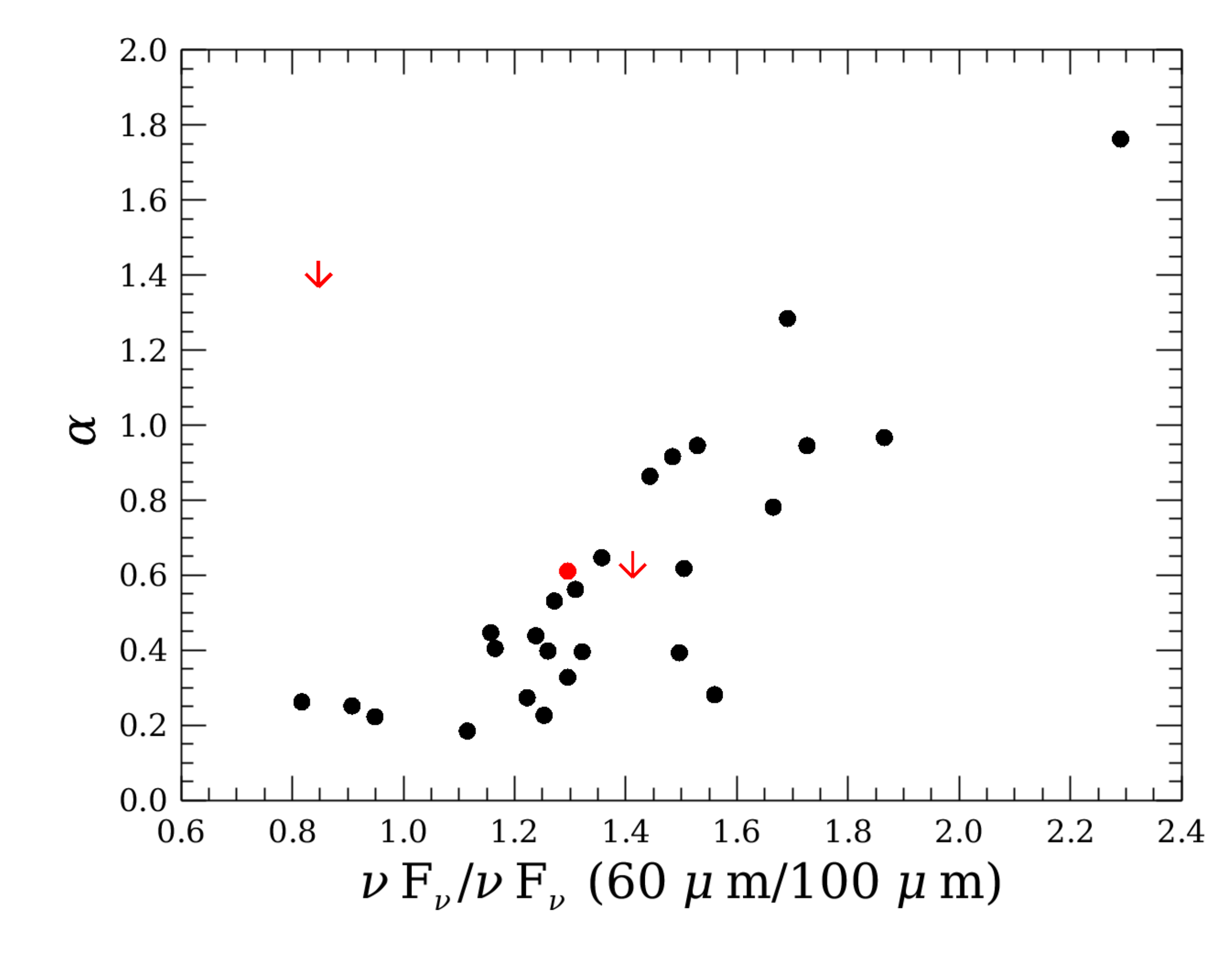} 
\caption{The excitation parameter $\alpha$ is plotted as a function of dust temperature.  Plotted in black are the galaxies from \citet{Rosenberg}.  We do not have enough detections to cite any statistics, but our galaxies (red) are consistent with the previous observations.}
\label{alpha_dust}
\end{figure}

\citet{Rosenberg} find that $\alpha$ has a positive correlation with dust temperature. This correlation suggests the presence of dense, warm molecular gas implies the presence of warm dust. Though we do not have enough detections to claim any trends, our sparse sample is consistent with their result, as seen in Figure \ref{alpha_dust}.

\subsection{\nii 205 \micron\ as a Star Formation Tracer}
In addition to some of the CO transitions being used to measure star formation in galaxies, the \nii 205 \micron\ line from the SPIRE spectra is being explored as a possible tracer of star formation. Nitrogen is ionized by UV emission from hot young stars with lifetimes on the order of a few to ten Myr, and thus provides a measure of the SFR on relatively short timescales, similar to the way in which thermal radio free-free emission or the optical Balmer lines are used as a star formation tracer.  

\citet{Zhao2013} find a tight correlation between IR luminosity and the \nii 205 \micron\ luminosity for a sample of LIRGs, with log(L$_{IR}$) = (4.51$\pm$0.32) + (0.95$\pm$0.05)log(L$_{[N II]}$) and a scatter of 0.26 dex in L$_{IR}$. \textbf{ We do not find such a relation among our ETGs with a similar analysis using FIR luminosity, but our sample size is only 9 galaxies, while \citet{Zhao2013} are working with a sample size of 70 galaxies, so more observations of ETGs could improve this relation.}  We also see a large variation in the relative strength of the \nii 205 \micron\ line and the mid J CO lines, a ratio which should be relatively constant assuming both scale with the amount of star formation.  The \nii 205/CO(5-4) ratio varies from 0.85-34, while the \nii 205/CO(6-5) and \nii 205/CO(7-6) ratios range from 0.83-44.4 and 0.75-15.5, respectively.  Thus, the \nii 205 \micron\ line does not appear to scale with the mid J CO lines.  

\begin{figure}
\epsscale{1.0 } 
\plotone{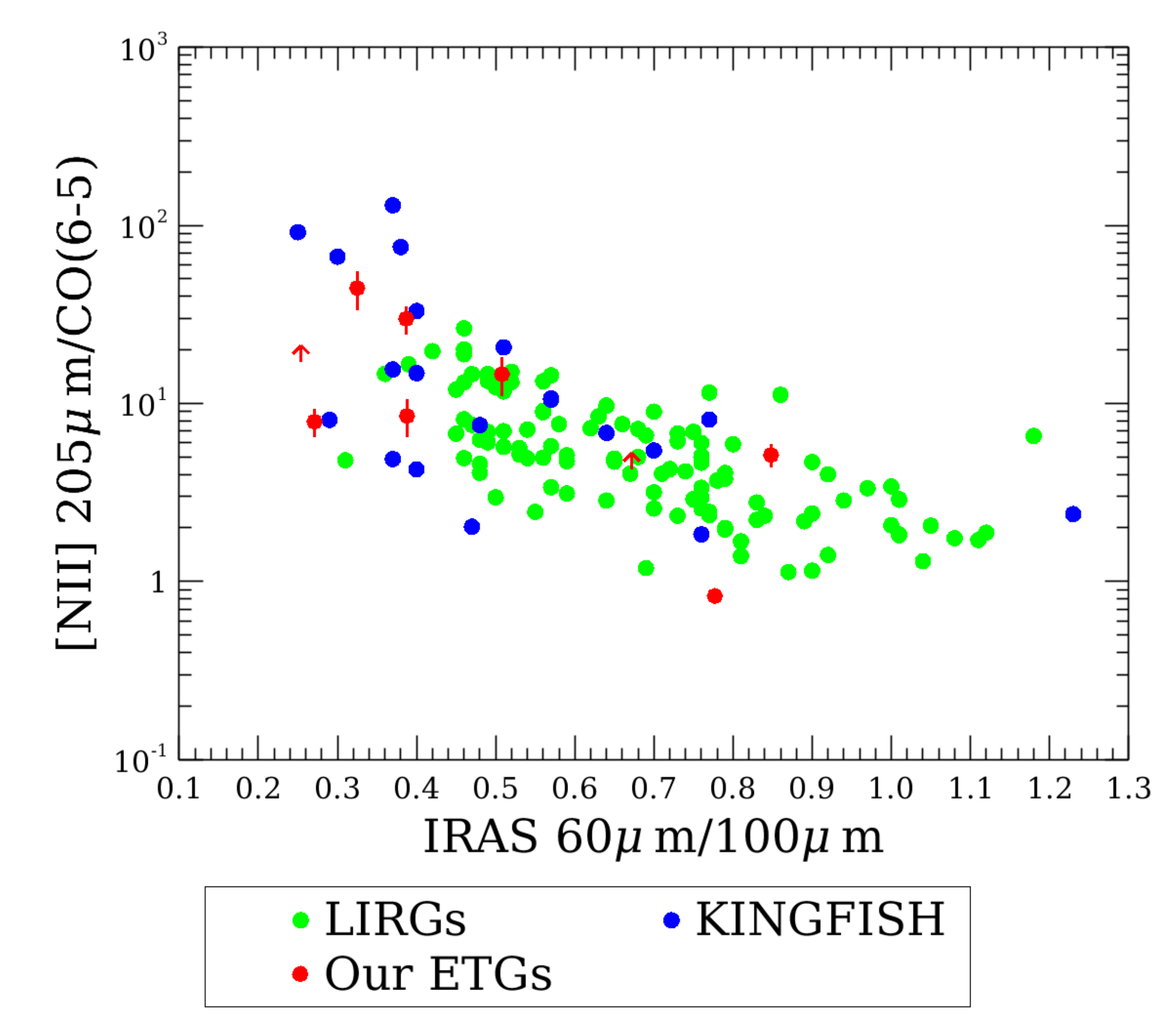} 
\caption{The \nii 205/CO(6-5) ratio is plotted against a proxy for dust temperature.  Green points are LIRGs from \citet{Lu2017}, while blue points are galaxies from the KINGFISH survey \citep{Kenn2011, Kam2016}.  Our ETGs are plotted in red.  There is a tendency for the ratio to decrease as the dust temperature increases. }
\label{NIICO}
\end{figure}

However, \citet{Lu} find a trend for the \nii 205/CO(7-6) ratio to decrease with increasing dust temperature.  Their results, along with discussion in \citet{Zhao2013}, imply a deficit in the \nii 205 \micron\ line at higher dust temperatures.  Because we have more detections of the CO(6-5) line, we have recreated the middle panel of Figure 2 from \citet{Lu} with the \nii/CO(6-5) ratio using published data from \citet{Lu2017}, along with data for galaxies from the KINGFISH and \textit{Beyond the Peak} surveys published in \citet{Kam2016}.  Our ETGs and the spiral galaxies appear to continue the trend seen in the LIRGs, as seen in Figure \ref{NIICO}.  Thus their claim that the \nii/CO line ratio can infer dust temperature, and hence the \textbf{intensity} of the ISRF, is strengthened.  This method would be useful for situations when dust continuum detections are unavailable, such as in high redshift galaxies.

\subsection{Neutral Carbon}
\label{sec: carbon}
In the standard onion skin PDR model, the deepest layers of a PDR only contain carbon in the form of CO. Neutral carbon exists alongside molecular hydrogen in a shell surrounding the CO core, and finally there is a layer of C+ on the exterior surface of the PDR. Since the layers of gas outside of the core cannot be traced by CO, the gas is deemed `CO dark'.  However, observational evidence is mounting that neutral carbon is not relegated to a simple layer outside of the CO, but exists alongside CO in the interior of molecular clouds.  A newer model of clouds composed of PDR `clumps' has been introduced to explain these observations.

\ci (1-0) emission is found to have similar spatial and velocity structure to $^{13}$CO emission in the Orion giant molecular cloud (GMC) \citep{Ikeda}.  \citet{Ikeda} also derive the ratio of \ci/CO column densities, and find that it varies from 0.2-2.9 on the edges of molecular clouds, while it is roughly constant (0.1-0.2) in the cloud interiors, suggesting CO and \ci are coexistent within molecular clouds. They find \ci column densities of a few to ten times 10$^{17}$ cm$^{-2}$. The nearly constant ratio between CO and \ci  in the cloud interiors lends credence to the clumpy PDR model, and slight discrepancies in the ratio from region to region are postulated to be the result of differing cloud size distributions. \citet{Ikeda} put forth an argument to constrain clump sizes between 0.03 and 0.25 pc based on the \ci/CO column density ratio, with larger clumps corresponding to lower ratios.  However, observations of  \ci with higher spatial and velocity resolution will be necessary to affirm this hypothesis.  

The spatial correlation found by \citet{Ikeda} could be explained by low density PDRs where \ci abundance is expected to be greater, or by FUV photons penetrating deep into molecular clouds and triggering emission from \ci coating the many clumps comprising a cloud, but the intensity correlation shown in \citet{Keene1997} along with similar temperatures derived from CO (and other means) and \ci by \citet{Stutzki} strongly suggest that \ci and CO reside in the same molecular gas clouds and trace the same molecular hydrogen.

At low temperatures the structure of neutral carbon effectively has 3 accessible energy levels.  There are three fine-structure levels in the 3P ground state, where the \ci(2-1) and (1-0) transitions arise.  Because of the simple structure, the fundamental (1-0) emission line is a good tracer of the column density as it does not have a strong dependence on excitation temperature.  On the other hand, the \ci(2-1) transition is quite sensitive to the temperature, making the ratio of the \ci lines a good tracer of the gas conditions.  

\subsubsection{Carbon as a Temperature Probe}
\label{sec: temperature}
The two emission lines of neutral carbon found in the SPIRE spectra can be used to determine kinetic temperature of the gas by utilizing the fact that the excitation temperature is equivalent to the kinetic temperature assuming local thermodynamic equilibrium (LTE).  In a survey of 17 nearby LIRGs, \citet{Kam2014} find a similar range of temperatures (20-40 K) as \citet{Walter} find for a sample of high redshift galaxies, and conclude that \ci is not particularly useful for determining the excitation conditions of the gas, as it returns similar values across a wide spread of distance and luminosity. \citet{CarilliWalter} conclude that \ci is probably measuring neutral atomic gas and its properties are not strongly affected by star formation activity.  We computed the excitation temperature for our sample of galaxies using the \ci line ratio and find a range of temperatures of $\sim$12-25 K, with some being lower limits. The ETGs are roughly consistent with the values found in LIRGs and high-redshift galaxies, though they tend to be lower, as there are two measurements below 20 K, and several others possibly as low as 12-15 K.  In Section \ref{sec:CICOSFR} we compare these temperatures to those found using RADEX models for the ratios of the \ci lines to their neighboring CO transitions.

\subsubsection{Measuring Gas Mass with \ci}
\label{sec: gasmass}

The \ci lines can be used to derive the global molecular gas mass, a concept used by \citet{PapaGreve}, who use just the J=1-0 transition, and shown again by \citet{Kam2014} who use both J=1-0 and J=2-1 and Boltzmann population distributions to determine the total molecular gas mass. \textbf{Additionally, a nearly linear relation between CO(1-0) luminosity and both \ci(1-0) and (2-1) luminosities was found by \citet{Jiao}. } The largest uncertainties when using \ci as a tracer of molecular gas stem from the ratio of the number density of neutral carbon atoms to the number density of H$_2$ molecules. \citet{PTV} point out many factors that affect the C/CO abundance ratio, including turbulence and the effect of cosmic rays, however they point out these same processes affect the X$_{CO}$ factor, and it is still widely used and approximately constant.  The small range of N(CI)/N(CO) of 0.1-0.2 found by \citet{Ikeda} supports the idea that the X$_{CI}$ factor may be approximately constant, too, but large surveys in the future will be necessary to establish an X$_{CI}$ value the community can trust. 

\begin{figure}
\epsscale{1.0 } 
\plotone{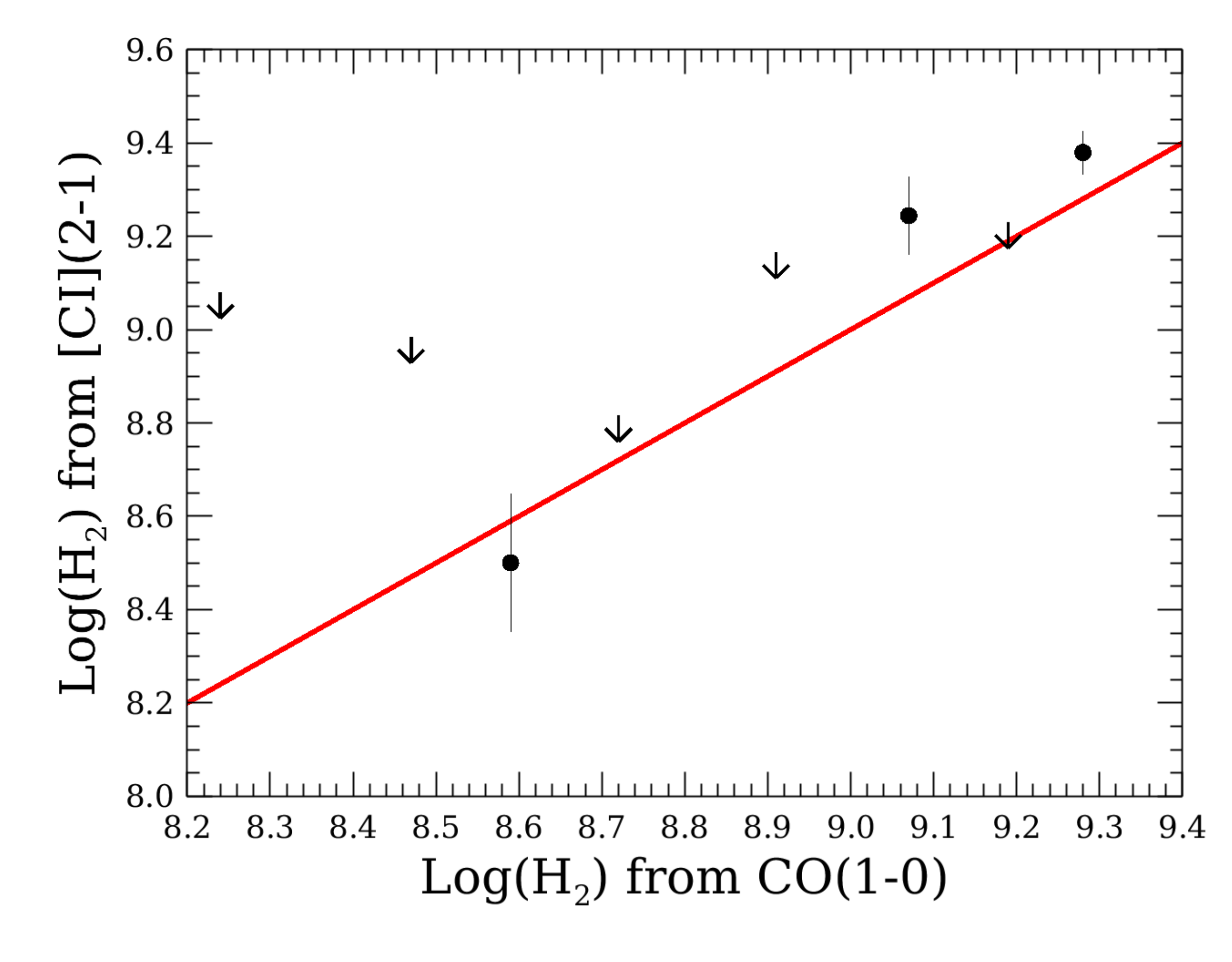} 
\caption{Molecular gas masses derived from both \ci emission lines are compared to masses derived from CO(1-0). The masses derived from \ci agree fairly well with masses derived from CO.  The red lines indicate a one to one correspondence. The CO masses are estimated using the ratio N(H$_2$)/I$_{1-0}$ = 3$\cdot$10$^{20}$ cm$^{-2}$ (K km s$^{-1}$)$^{-1}$ and the corresponding integrated equation from \citet{Young2011}.  The molecular gas masses are in units of M$_{\sun}$. Uncertainties of the masses derived from CO are typically around 20\% (0.08 dex). }
\label{gasmass}
\end{figure}

\citet{PapaGreve} use an abundance ratio of 3$\cdot$10$^{-5}$ C/H$_2$, which is roughly the average between the value derived in \citet{Ikeda} from \ci and CO observations of the Orion A and B clouds and the value derived in \citet{White1994} for the starbursting nucleus of M82. \citet{Kam2014} assume an equivalency between masses derived from CO(1-0) and \ci, and use the relation to calculate an abundance ratio of 10$^{+20}_{-7}\cdot$10$^{-5}$ C/H$_2$.  Since the starbursting nucleus of M82 does not represent ETGs well, we adopt the value in \citet{Ikeda} of 10$^{-5}$, and have carried out calculations based on Boltzmann populations as in \citet{Kam2014} using the pair of \ci transitions, and have estimates of the total molecular gas mass for each galaxy. They are compared to masses derived from CO(1-0) in Figure \ref{gasmass}.  

For our small sample there is \textbf{good agreement} with the mass derived from CO. In galaxies with lower metallicities or higher UV radiation fields where CO is less reliable as a mass tracer, \ci could be quite useful in providing a more accurate measure of the molecular gas content.  Both of these conditions would be applicable to high redshift galaxies, and the \ci lines may be shifted to frequencies observable from the ground in high redshift sources. Had we used the conversion factor derived in \citet{Kam2014}, it would have given H$_2$ masses significantly lower than those inferred from CO.  \textbf{This indicates a difference between ETGs and the sources (most of which are luminous infrared galaxies) in \citet{Kam2014}, in that the ETGs have somewhat weaker \ci emission relative to CO.  Further work should be done to explore whether this method extends to lower metallicity objects, as well as what differences exist between various galactic morphologies. }

\subsection{Gas Cooling Budgets}
\label{sec:budget}
The ratio of \cii/FIR or (\cii + \oi)/FIR is often used as an estimate for the heating efficiency of gas by photoelectrons off of dust grains. Dust is heated by UV photons and then emits infrared photons, while photoelectrons ejected off of these dust grains heat the gas, which in turn is cooled via spectral emission lines.  By examining these ratios one can determine how large of a role these lines play in cooling the gas, as well as how that role changes in different conditions. For instance, as the dust temperature increases, the \cii line becomes less efficient at  cooling the gas and the \cii/FIR ratio decreases.  Another result of \citet{Rosenberg} is that the \ci line experiences a similar deficit as the \cii line as a fraction of FIR, and they interpret this as star formation that is dominated by ultracompact \hii regions. In such a region, the dust would absorb a larger fraction of the UV photons, driving down the relative flux from all emission lines.  Conversely, CO does not show a CO/FIR deficit.  This could be because the CO molecules reside in a different phase of the ISM than C and C+.   Additionally, CO is responsible for much more of the gas cooling than PDR models predict, reaching values of tens of percent, while models only predict 3-5\% of the total gas cooling budget to be attributed to CO \citep{Meijerink}.  \citet{Rosenberg} surmise that the CO heating mechanism is different, and is not one that leads to dissociation or ionization.  There is not a deficit in our ETGs for CO/FIR nor \ci/FIR, though much of our data consists of upper limits.  The values of ratios are consistent with the values for other galactic morphologies.

\citet{Gerin} find a trend of \cii/\ci(1-0) to decrease as \ci(1-0)/FIR increases, however the 2 ULIRGs in their sample do not fit the trend.  They also find a correlation between \cii/FIR and \ci(1-0)/FIR, barring Arp 220 and Mrk 231 which fall below the trend line.  The interpretation they give is that these correlations are the result of changes in the value of G$_0$ (the incident far ultraviolet (FUV) flux of photons between 6 and 13.6 eV in units of 1.6$\cdot$10$^{-3}$ ergs cm$^{-2}$ s$^{-1}$, the local ISRF). Around G$_0$=10-100, the efficiency of the gas heating reaches a maximum, which saturates the \ci(1-0)/FIR or \cii/FIR ratios.  If the value of G$_0$ increases from this point, the gas heating will be less efficient, while the dust is still being heated the same amount per photon.  PDR models show the FIR flux to increase linearly with G$_0$ beyond this range, while \cii increases logarithmically and \ci(1-0) stays relatively constant, leading to \ci(1-0)/FIR and \cii/FIR ratios that decrease as the radiation field increases.  Since \ci(1-0) has a weak dependence on G$_0$ and n, the \cii/\ci(1-0) will increase with stronger radiation fields.  Their PDR models find that the majority of the galaxies studied lie within the parameter space G$_0$=10$^{2}$-10$^{3}$, and n=10$^{2}$-10$^{5}$ cm$^{-3}$.  

\begin{figure}
\epsscale{1.0 } 
\plotone{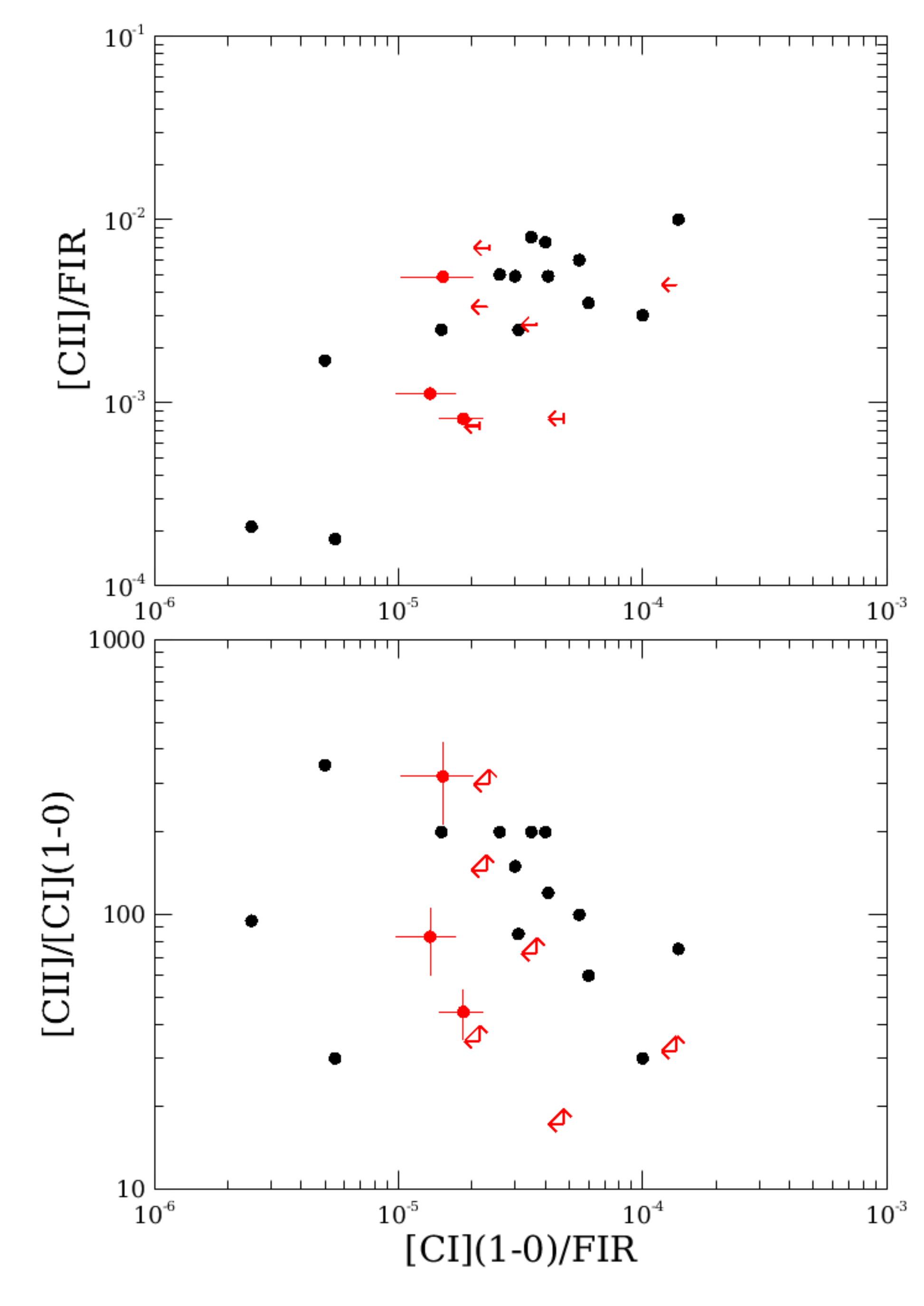} 
\caption{The \cii/FIR and \cii/\ci(1-0) ratios plotted against \ci(1-0)/FIR. Our ETGs are plotted in red, while the galaxies from \citet{Gerin} are plotted in black. We find the same increasing trend for \cii/FIR as \citet{Gerin}, though not at a statistically significant level, and the same is true for the decreasing trend for \cii/\ci(1-0) that they find for their galaxies.  However, our galaxies do not span as wide of a range in \ci(1-0)/FIR as their sample does.}
\label{CarbonRatios}
\end{figure}

We have plotted the same ratios for our galaxies, shown in Figure \ref{CarbonRatios}. We find a range of \ci(1-0)/FIR values for ETGs of around 10$^{-5}$ to 4.2$\cdot$10$^{-5}$. This is more narrow than the range of \ci(1-0)/FIR found for spirals or irregulars/mergers by \citet{Gerin}, but falls within their bounds.  They find values of \ci(1-0)/FIR between 10$^{-5}$ and 10$^{-4}$ for spirals, and 2$\cdot$10$^{-6}$ to 2$\cdot$10$^{-4}$ for irregulars/mergers.  Our galaxies show the same tendency for \cii/FIR to increase with \ci(1-0)/FIR, though not at a statistically significant level using a Kendall rank test. The same is true for the decreasing trend of \cii/\ci(1-0) that they find, albeit over a much more narrow range of \ci(1-0)/FIR than that spanned by their sample.  The same plots were made using the \ci(2-1) emission line, as well, with similar results. The \cii/FIR ratio increases with \ci(2-1)/FIR, but with even less significance, and the \cii/\ci(2-1) ratio decreases with respect to \ci(2-1)/FIR. We find a range of \ci(2-1)/FIR values of 10$^{-5}$ to 3.8$\cdot$10$^{-5}$. 


\subsection{PDR Models}
\label{sec:pdrt}
To learn more about the conditions of the gas in our galaxy sample, we utilized the online PDR toolbox (PDRT), a set of PDR models which take a given set of gas phase elemental abundances and grain properties and solves for the equilibrium chemistry, thermal balance and radiation transfer through a PDR layer for a fixed hydrogen nucleus density, n, and incident FUV intensity, G$_0$.  Observed line ratios can be compared to their theoretical counterparts to infer the possible combinations of n and G$_0$ capable of producing the observations.  Solutions are then found through a $\chi^{2}$ minimization routine applied to the line ratios computed from the input fluxes  \citep{KWH06, PW08}.  Although the ISM has a fractal structure, such that no single density or ISRF can accurately characterize the ISM of a galaxy, these volume averaged quantities can help us understand the differences from one galaxy (or morphology) to the next.  For future work it would be useful to model various density distributions within a beam to explore their effects on PDR models.

To a certain degree, you get out of a PDR model what you put in.  For example, using emission lines that only originate in dense gas such as high J CO lines will likely return a solution with a high density. One approach to this issue is to explicitly consider different phases of the ISM (molecular clouds, diffuse warm neutral gas, etc) separately, as in \citet{Abdullah}.  But our early-type galaxies have significantly worse spatial resolution and lack much of the ancillary data needed to characterize multiple phases of the ISM.  Hence we take a different approach.  We effectively lump all the phases of the ISM together, but differences in the global average quantities may still illuminate differences in the mix of phases from one galaxy to the next. To explore this conundrum, we have used every possible combination of inputs for the PDRT (with a minimum of 3 and a maximum of 6 spectral lines or the FIR continuum).  We use combinations of the pair of \ci lines, \cii 158, \oi 63, CO(1-0), and FIR for galaxies with SPIRE observations, and only \cii 158, \oi 63, CO(1-0), and FIR for our ETGs without SPIRE data. This subset of lines spans the ionized, neutral, and molecular components of PDRs, but doesn't include warm molecular gas that the mid and high J CO lines originate from.  \citet{Mal01} use only the \cii 158, \oi 63, and FIR continuum, so we also find values using only \cii, \oi, and FIR for a direct comparison to the values derived by \citet{Mal01}. The \cii and \oi fluxes for our ETGs are taken from \citet{Lapham}.

\begin{figure*}
\hbox{
\includegraphics[scale=0.4,trim=10mm 5mm 0mm 0mm]{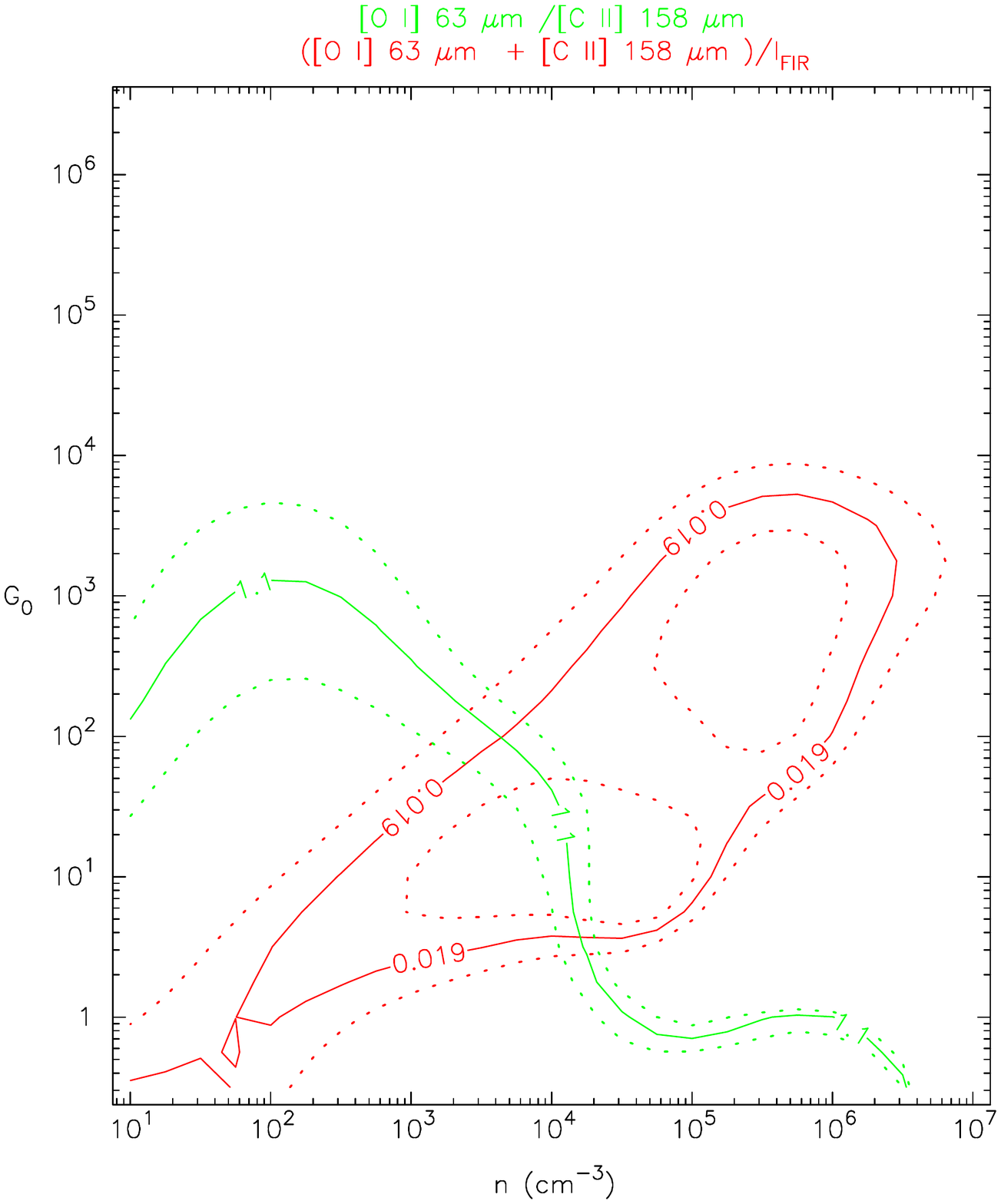}
\includegraphics[scale=0.4,trim=10mm 5mm 0mm 0mm]{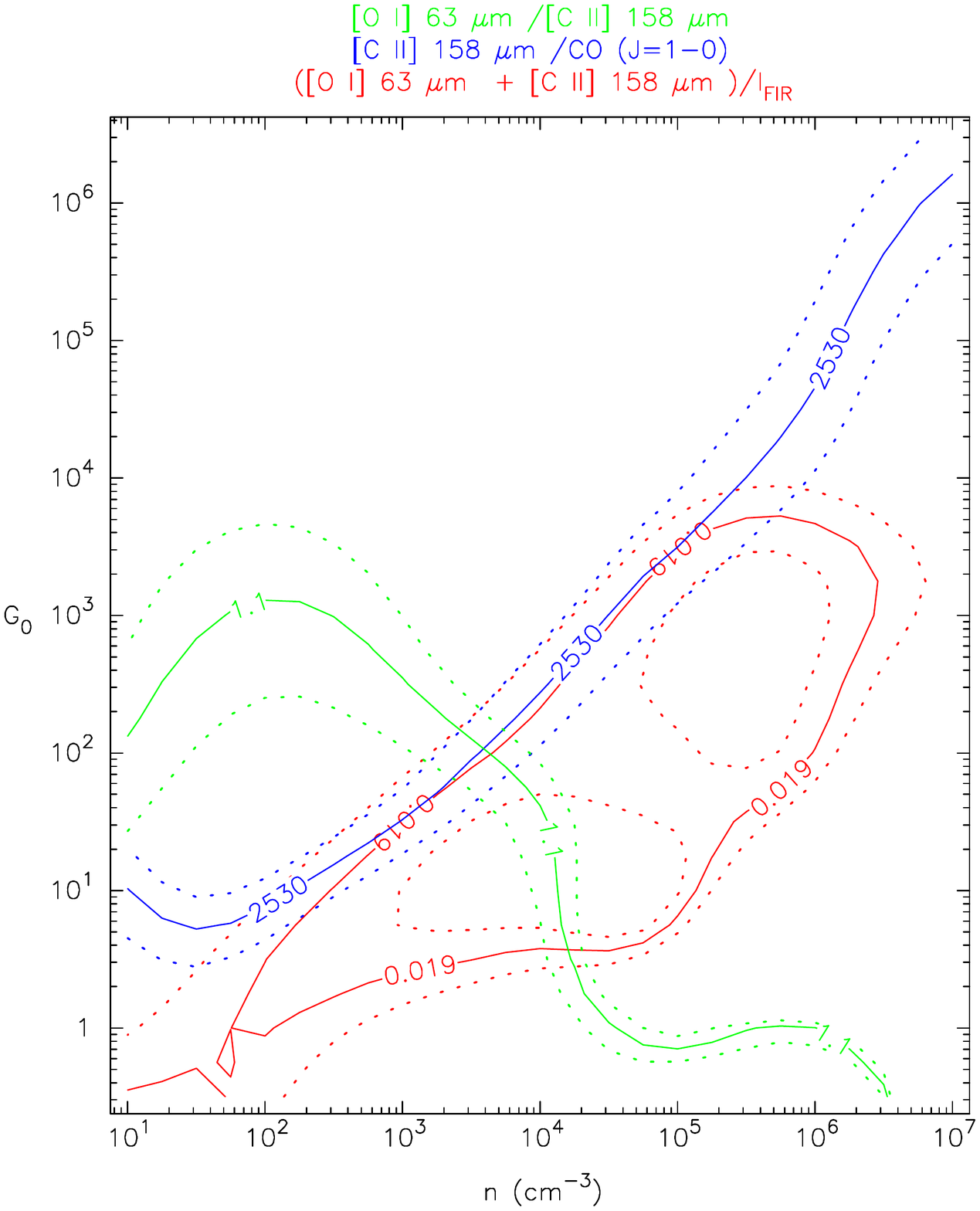}
}
\caption{Left: PDRT solution contours for IC 1024 using only \cii, \oi, and FIR. Right: PDRT solution contours for IC 1024 using \cii, \oi, FIR, and CO(1-0).  Uncertainties in the ratios are indicated by dashed lines. For this galaxy the addition of the CO(1-0) emission line reveals an obvious solution.}
\label{PDRT_good}
\end{figure*}

\begin{figure*}
\hbox{
\includegraphics[scale=0.4, trim=10mm 5mm 0mm 0mm]{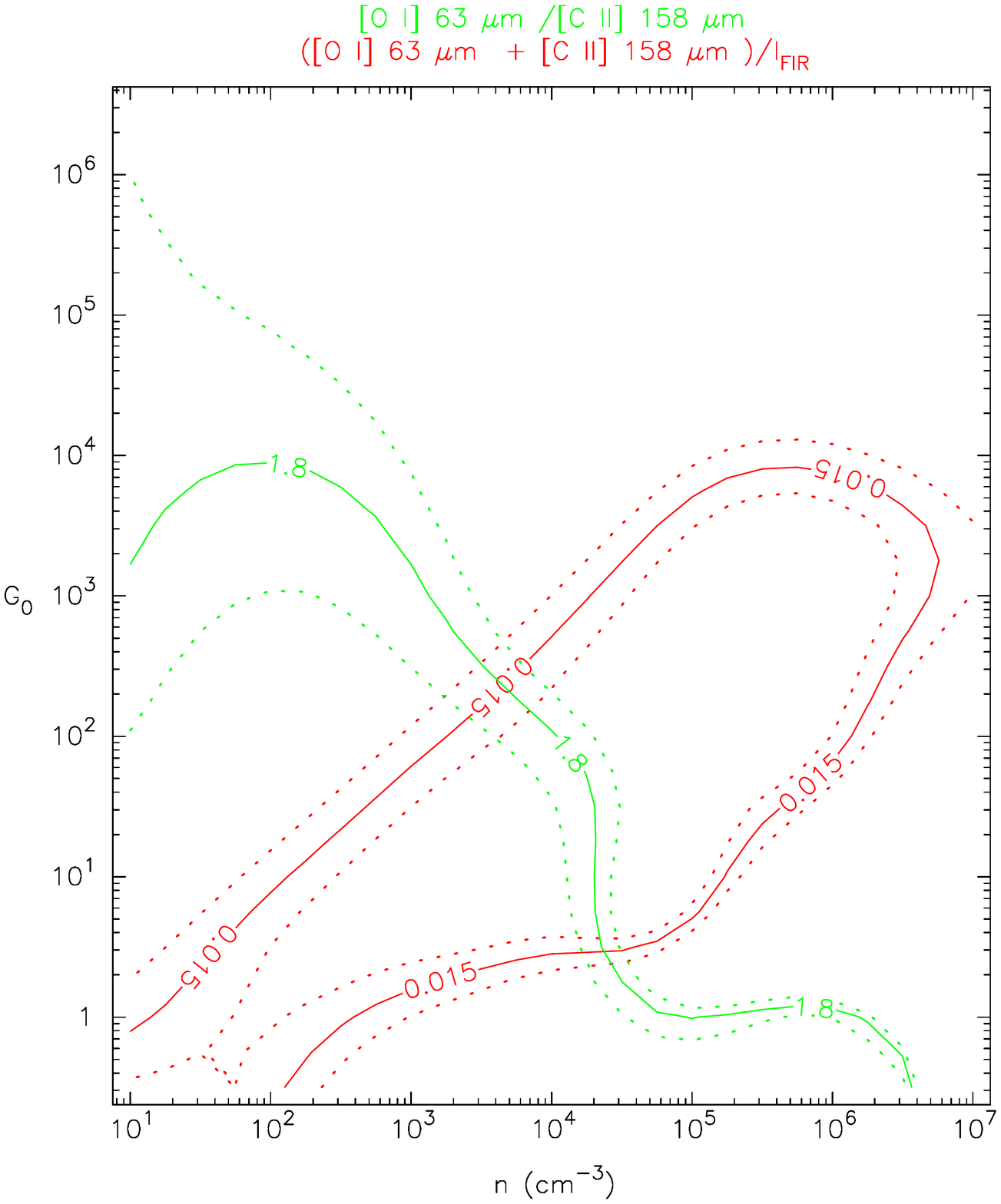}
\includegraphics[scale=0.4, trim=10mm 5mm 0mm 0mm]{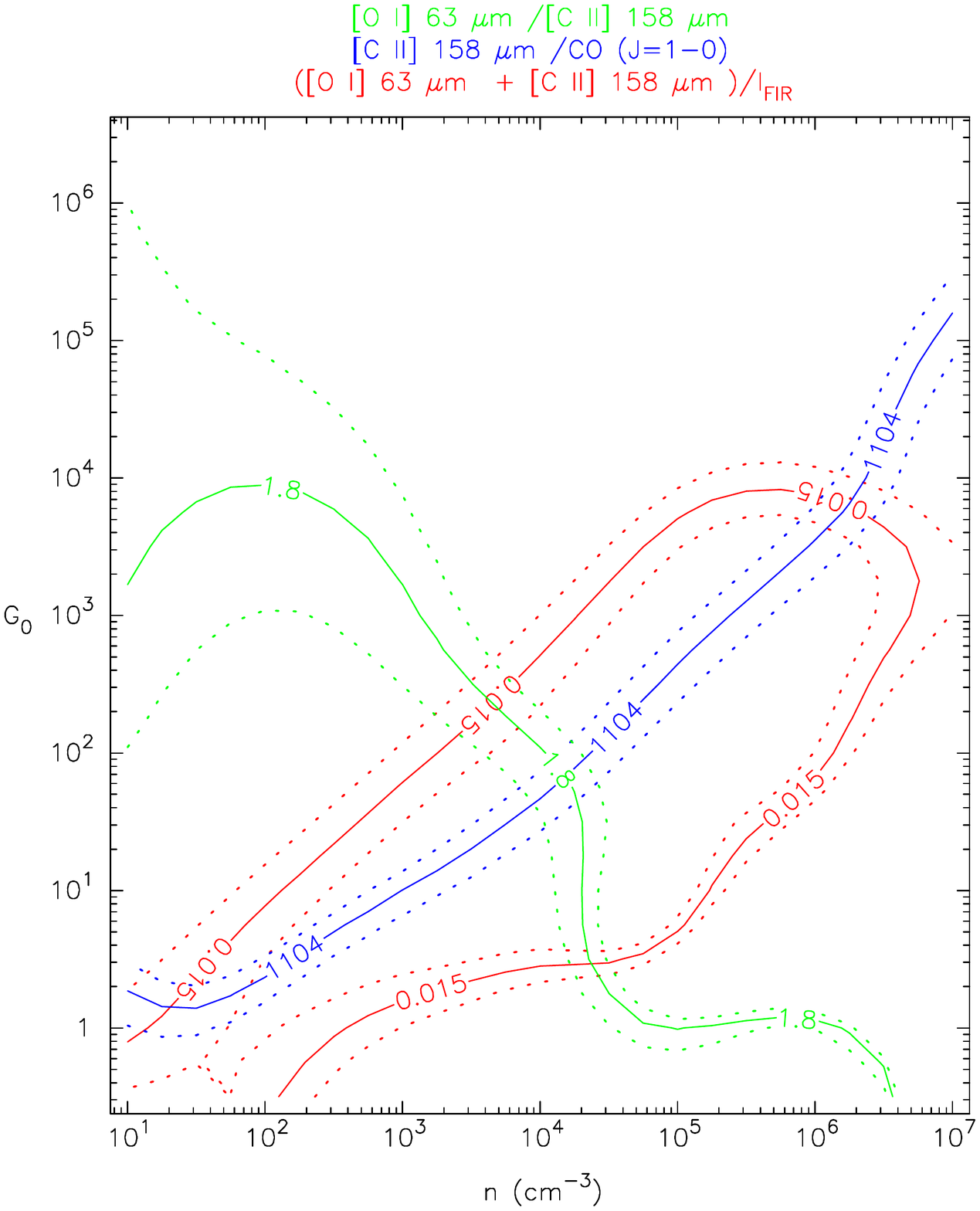}
}
\caption{Left: PDRT solution contours for IC 0676 using only \cii, \oi, and FIR. Right: PDRT solution contours for IC 0676 using \cii, \oi, FIR, and CO(1-0).  Uncertainties in the ratios are indicated by dashed lines.  For this galaxy the addition of the CO(1-0) emission line creates two more possible solutions.}
\label{PDRT_bad}
\end{figure*}

Once a set of input fluxes has been chosen, contour plots of the G$_0$ and n values that reproduce the observed flux ratios are overlaid, and their intersections indicate potential solutions.  A pair of contours typically has more than one solution, so additional contours can help break this degeneracy and find a favorable solution, as shown in Figure \ref{PDRT_good}.  However, the addition of more contours can also produce more potential solutions, as shown in Figure \ref{PDRT_bad}.  In either case, the final solution is based on a $\chi^{2}$ distribution for every flux ratio. The solutions for each galaxy can be found in Table \ref{PDRT_table}.

\subsubsection{PDR Models Using Only \cii and \oi Lines}
\label{PDRCIIOI}
Following the recommendations by \citet{KWH99}, we divide the observed FIR flux by a factor of two.  This is because the model assumes emission only comes from the side of the cloud illuminated by the UV radiation, when it likely comes from both near and far sides of the cloud as the cloud is optically thin to dust continuum IR photons.  Additionally, \citet{KWH99} recommend to increase the \oi emission by a factor of two.  They argue that because the line is optically thick, the variety of different PDR orientations will lead to only half of the emitted \oi flux being observed.  The extreme cases would be a cloud lit from behind or face on.  No emission would be observed in the former and all of the emission would be observed in the latter.  The \cii line also needs a correction since part of the total flux comes from \hii regions rather than PDRs.  This contribution can be quantified if observations of the \nii 122 and 205 \micron\ lines are available. Galaxies with SPIRE observations use corrected \cii fluxes (flux only from PDR contributions) determined in \citet{Lapham}, otherwise a flat 30\% is removed for ionized gas contributions.  These corrections are applied to all of our galaxies, as well as the comparison sample from \citet{Mal01} in the following analysis.

\begin{figure}
\epsscale{1.0 } 
\plotone{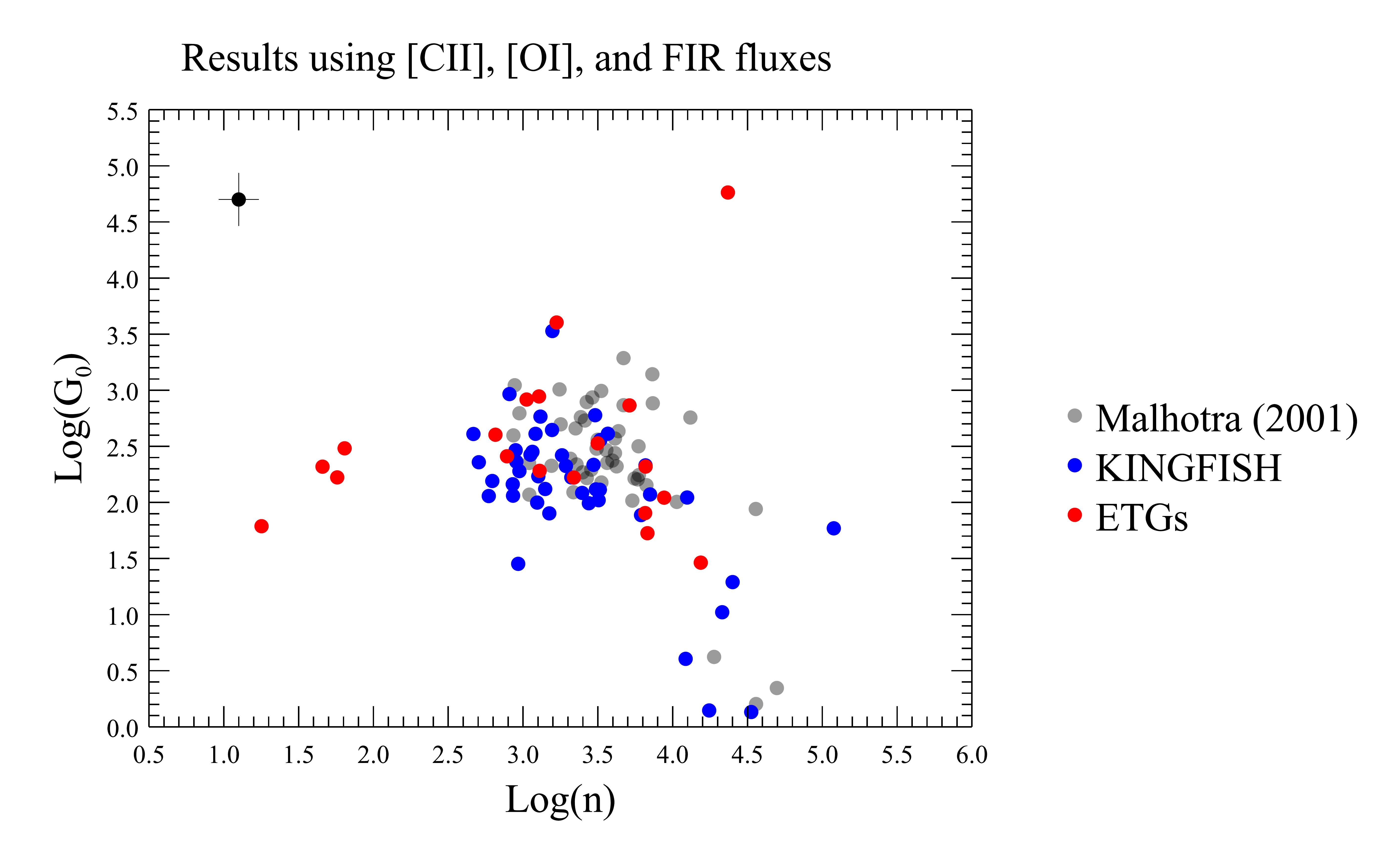} 
\caption{Results from the PDR toolbox using corrected \cii, \oi, and FIR fluxes. Our ETGs are plotted in red, while galaxies from the KINGFISH sample are plotted in blue. The grey data points are galaxies from \citet{Mal01}.  The black point in the upper left represents typical uncertainties. }
\label{PDRT_corrected}
\end{figure}

\textbf{When we relegate our models to only the \cii, \oi, and corrected FIR fluxes, five ETGs fall along a line above the comparison samples (NGC 3607, 4429, 4459, 4526, and 5866), while the remaining fourteen drift towards higher density, lower UV intensity solutions where the comparison galaxies lie.  This can be seen in Figure \ref{PDRT_corrected}.  ETGs having lower ISRF values would be consistent with their older stellar populations, however, the galaxies with the lowest UV intensity solutions are from the KINGFISH and \citet{Mal01} samples.  Thus the majority our ETGs are similar to spiral galaxies, while some exhibit different behavior. The (\cii + \oi)/FIR ratio is likely what drives the unusual behavior of a handful of our ETGs, as shown in Figure \ref{OCFIR_contour}.  }


\subsubsection{PDR Models Including CO and \ci}
\label{PDRCOCI}
After modeling all of the allowed combinations of PDR lines chosen, the median values for n and G$_0$ for our ETGs are plotted \textbf{against the results from the  \citet{Mal01} sample found in the previous section (using only \ci, \oi, and FIR) in Figure \ref{PDRT_SPIRE}.}  Typical uncertainties for G$_0$ and n are also calculated from the various allowed flux ratio combinations, and are scaled to the uncertainty of the median by a factor of 1.25 (the square root of $\pi$/2).  \textbf{Our ETGs occupy a similar range as the spirals from the \citet{Mal01} sample.  NGC 4526 lies furthest from the group, with the highest density and a low value of G$_0$.}  Curiously, NGC 4526 was also found to have the largest \nii 122/\cii 158 ratio in \citet{Lapham}. Though these results may not be related, it could be an indication that metallicity plays a role in regulating the ISRF, a possibility that will be explored in future work.  It also appears that the galaxies known to be merger-driven starbursts (i.e. NGC 1222 and NGC 7465) or blue in color (NGC 2764) have relatively high values of G$_0$, while the quiescent Virgo Cluster members have lower values of G$_0$.  Galaxies with log($G_0$) $\geq$ 2 include NGC 1222, 2764, 3665, 4710, and 7465, as well as IC 1024 and IC 0676.  Galaxies below this threshold include Virgo Cluster members NGC 4429, 4435, 4459, 4526, 4694, as well as NGC 1266, NGC 3032, 3489, 3626, 4150, and NGC 5866, though NGC 1266 and 3626 are close with log(G$_0$) = 1.75.

The inclusion of the \ci and CO lines likely probes a denser phase of the ISM, possibly getting into the molecular cloud and not the PDR at the surface of the cloud.  Contour plots of G$_0$ and n show that solutions for \ci(2-1)/CO lie exclusively at large densities and low radiation fields for all of the galaxies in our sample as shown in figure \ref{CICO_contour}. A similar plot for the \cii/CO ratio does not restrict solutions to the same high n, low G$_0$ regime, seen in Figure \ref{CIICO_contour}.  Thus it appears the inclusion of the \ci emission lines is the primary driver for pushing the models to the high n, low G$_0$ solutions, however, it is not solely responsible, as the addition of only the CO(1-0) also pushes solutions to this regime as seen in Figure \ref{PDRT_SPIRE}.  We will revisit the \ci/CO ratios in Section \ref{sec:CICOSFR}, where RADEX models will be used to recreate the observed line ratios.

\begin{figure}
\epsscale{1.0 } 
\plotone{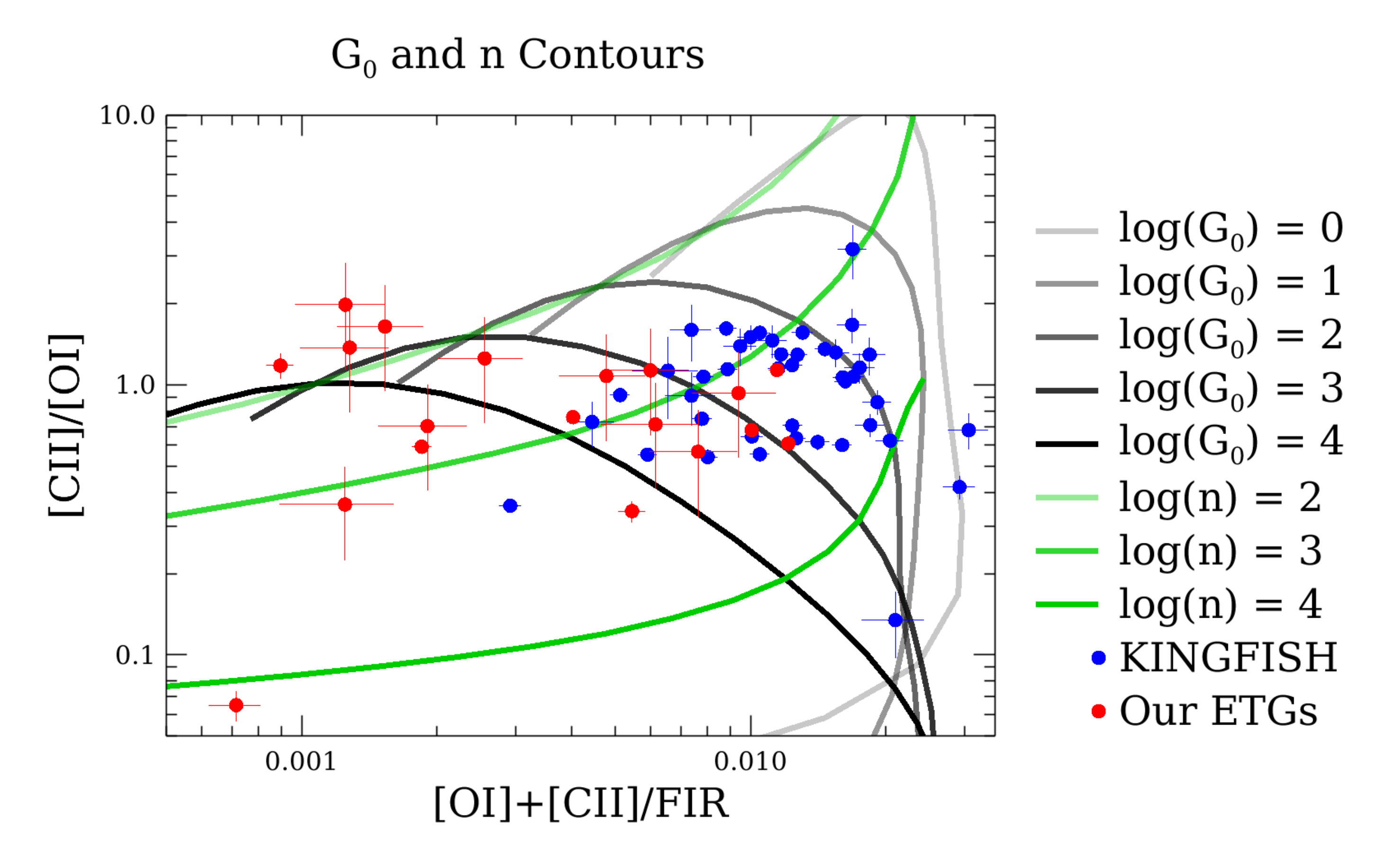} 
\caption{Contours for n and G$_0$ solutions corresponding to the \cii/\oi and (\cii+\oi)/FIR ratios.  Ratios for our ETGs are plotted in red, while ratios for galaxies from the KINGFISH sample are plotted in blue. NGC 4526 lies at a low \cii/\oi ratio because the majority (95\%) of its \cii emission comes from ionized gas.}
\label{OCFIR_contour}
\end{figure}

\begin{figure}
\epsscale{1.0 } 
\plotone{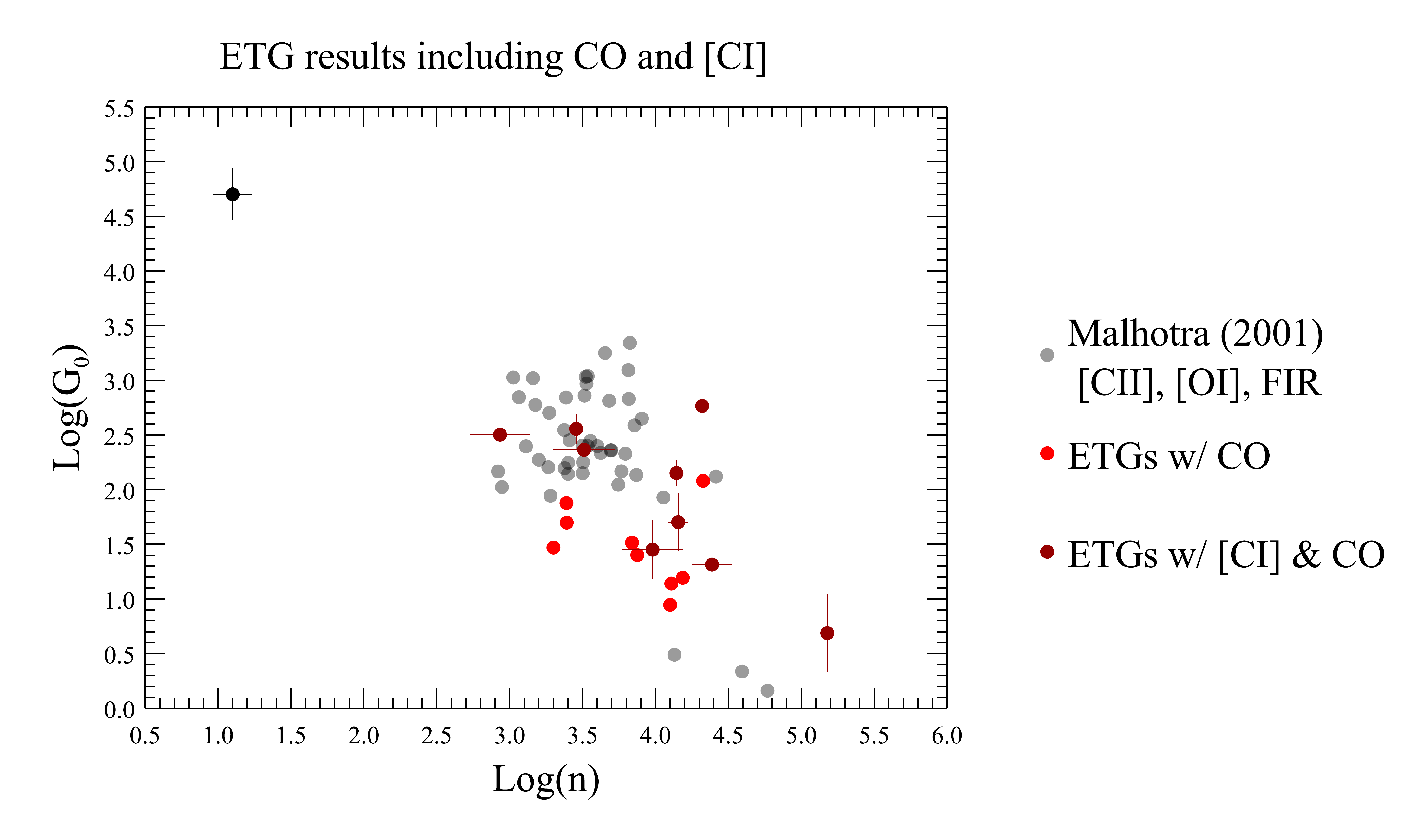} 
\caption{Results from the PDR toolbox using \cii, \oi, and FIR fluxes, as well as CO(1-0) and the pair of \ci emission lines. ETGs with only the additional CO(1-0) line are plotted in red, while the darker red represents the galaxies which make use of \ci lines from SPIRE as well.  The grey data points are galaxies from \citet{Mal01} computed from only \cii, \oi, and FIR fluxes. The black point in the upper left represents typical uncertainties.}
\label{PDRT_SPIRE}
\end{figure}

\begin{figure}
\epsscale{1.0 } 
\plotone{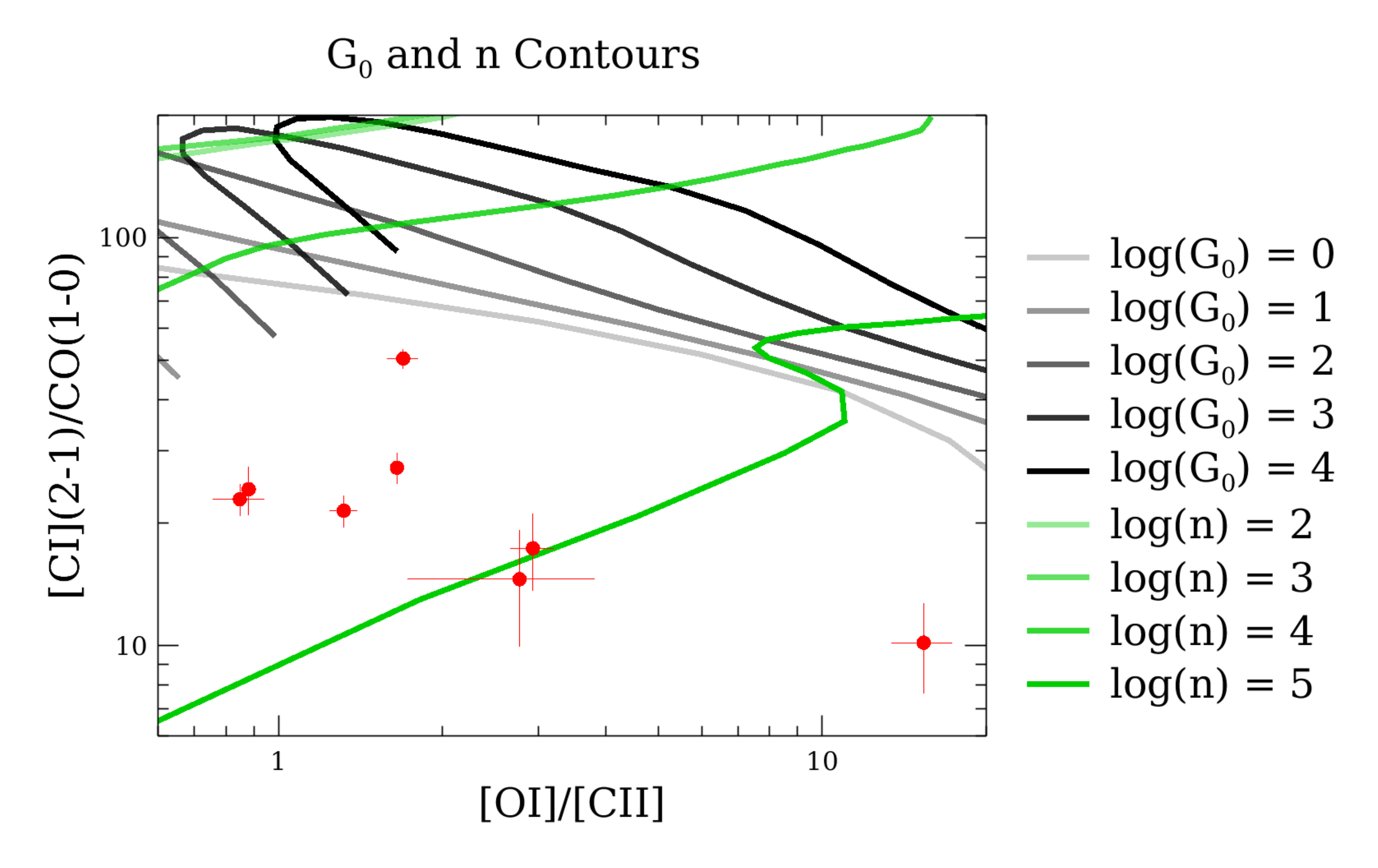} 
\caption{Contours for n and G$_0$ solutions corresponding to the \ci(2-1)/CO(1-0) and \oi/\cii ratios.  Ratios for our ETGs are plotted in red. NGC 4526 lies at a high \oi/\cii ratio because the majority (95\%) of its \cii emission comes from ionized gas.}
\label{CICO_contour}
\end{figure}

\begin{figure}
\epsscale{1.0 } 
\plotone{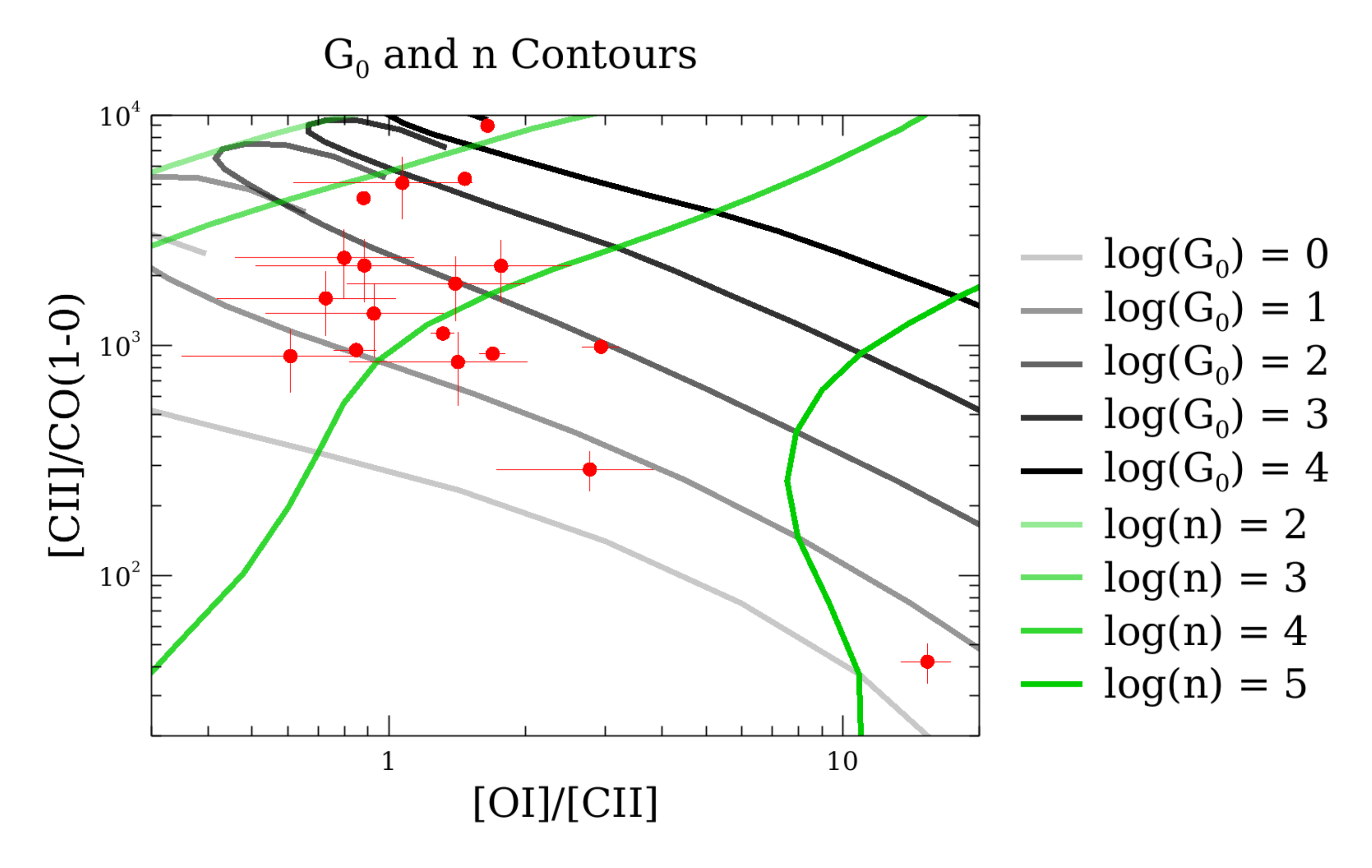} 
\caption{Contours for n and G$_0$ solutions corresponding to the \cii/CO(1-0) and \oi/\cii ratios.  Ratios for our ETGs are plotted in red. NGC 4526 lies at a high \oi/\cii ratio because the majority (95\%) of its \cii emission comes from ionized gas.}
\label{CIICO_contour}
\end{figure}

It is curious to note that values of G$_0$ and n which can reproduce the other line ratios (such as \oi/\cii and (\oi + \cii)/FIR) apparently cannot reproduce the observed \ci/CO line ratios.  Comparing Figures \ref{CICO_contour} and \ref{OCFIR_contour}, one can see the observed \ci/CO ratios disagree with the models.  The observed \ci flux is typically at least a factor of 5 fainter than what the models would predict based on \oi, \cii, and FIR.

\subsubsection{Density-UV Field Scaling Relations}
\label{Scaling}
\begin{figure}
\epsscale{1.0 } 
\plotone{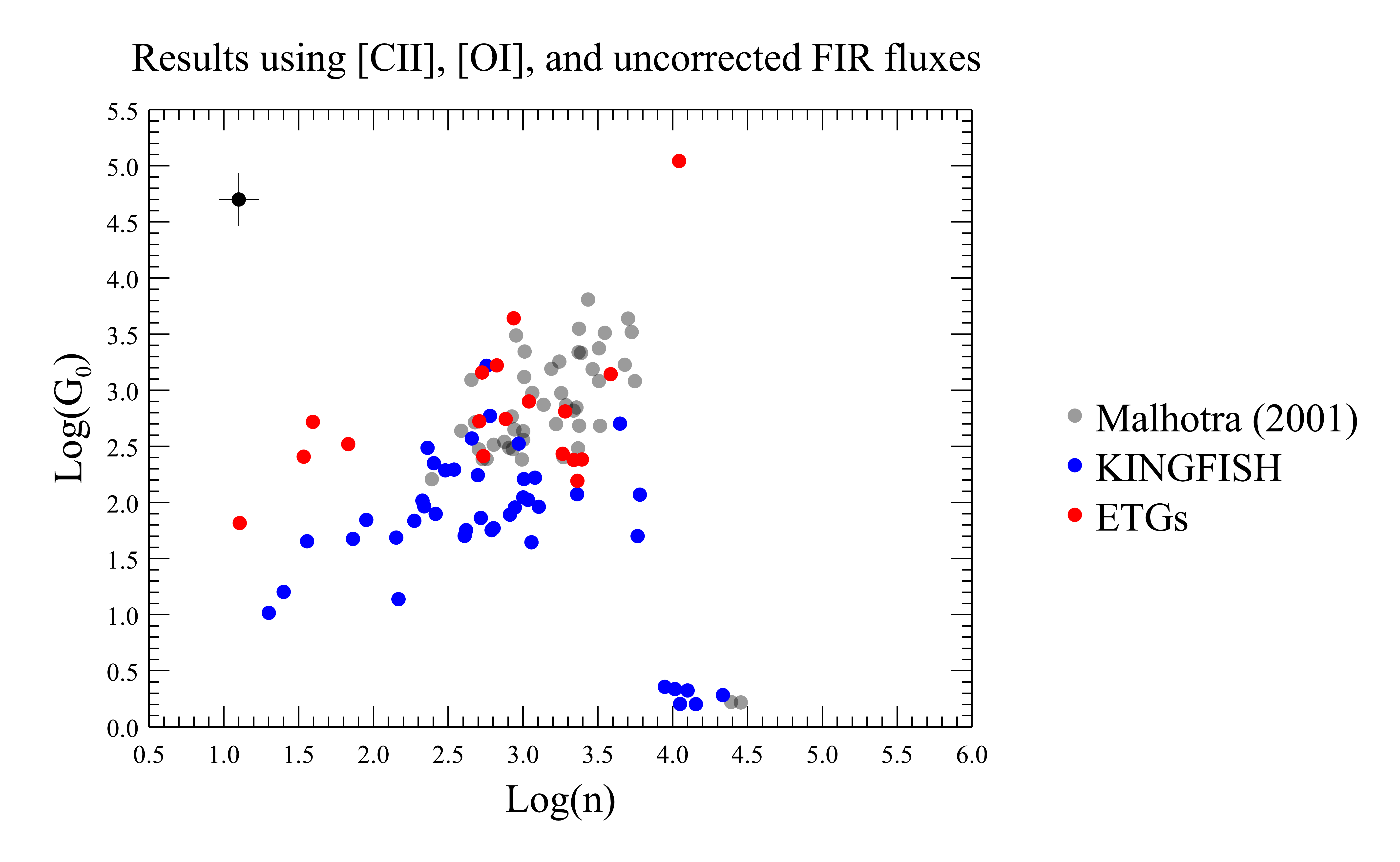} 
\caption{Results from the PDR toolbox using corrected \cii and \oi fluxes and uncorrected FIR fluxes. Our ETGs are plotted in red, while galaxies from the KINGFISH sample are plotted in blue. The grey data points are galaxies from \citet{Mal01}.  The black point in the upper left represents typical uncertainties.}
\label{PDRT_uncorrected}
\end{figure}

\textbf{One of the prominent results from the PDR modeling in \citet{Mal01} is a scaling relation between G$_0$ and $n$, with G$_0$ $\varpropto$ n$^{1.4}$.  This scaling relation was not reproduced by our PDRT models.  \citet{Mal01} use a similar method to correct the \cii flux with the \nii 122 \micron\ emission, and increase the \oi flux by a factor of two.  However, there is no mention of a correction to the FIR flux.  Thus, in an effort to reproduce their scaling relation we have also used the PDRT without correcting the FIR flux for and of the samples (our ETGs, the KINGFISH galaxies, and the galaxies from \citet{Mal01}).  These results are plotted in Figure \ref{PDRT_uncorrected}, and appear to be consistent with the G$_0$ $\varpropto$ n$^{1.4}$ trend found for the galaxies studied in \citet{Mal01}.  However, they still lie at lower values of the UV field strength G$_0$ than the original result in \citet{Mal01}.  }

There are two proposed explanations for the G$_0$ $\varpropto$ n$^{1.4}$ scaling relation observed in Figure \ref{PDRT_uncorrected} and \citet{Mal01}. The first is that it is the result of a Schmidt law on a smaller scale, with high gas densities leading to high star formation rates \citep{Schmidt}.  The exponential behavior agrees well with further work done on the Schmidt law by \citet{Kenn1998}, who find the global surface densities of star formation to scale with the surface density of gas by an exponent of 1.4. Another explanation offered by \citet{Mal01} is that the relation between G$_0$ and n is the result of geometry.  Assuming the FIR line and continuum emission is dominated by PDR contributions outside of young stars, Str\"{o}mgren sphere calculations predict the FUV flux at the neutral surface outside of the sphere to have a dependence of n$^{4/3}$. \citet{Mal01} prefer the Str\"{o}mgren sphere explanation.  First, they cite the high values of G$_0$ and n found by PDR models.  The large values of G$_0$ are only seen very close to O and B stars (around 1 pc), making it clear the results are probing local behavior, rather than global averages.  While some of this radiation escapes the molecular cloud and irradiates the diffuse ISM, the PDR model is weighted to the high G$_0$, high n component which is exactly where the Str\"{o}mgren sphere scaling is applicable.  Additionally, they have pressures derived from \oiii 88/\oiii 52 ratios that support this conclusion.

If we assume the latter explanation is what dictates the observed trend of G$_0$ $\varpropto$ n$^{1.4}$, it could potentially explain why our models with \ci or CO do not reproduce the trend.  The flux scaling predicted by Str\"{o}mgren spheres occurs at the surface illuminated by FUV photons, while the \ci and CO reside deeper in the cloud where the ISRF is weaker and the gas is more dense, leading to higher density, lower G$_0$ solutions.  However, this does not explain the galaxies that deviate from the trend even when only the \cii, \oi, and FIR fluxes are used in the PDR models.  For future work it would be useful to compute G$_0$ directly to test the accuracy of the PDRT solutions.

\subsection{Classifying Gas Conditions with \ci/CO ratios}
\label{sec:CICOSFR}
Line ratios of the \ci transitions to their neighboring CO transitions can be used to constrain the conditions of molecular gas in their host galaxy.  The \ci (1-0) emission line lies near the CO (4-3) emission line, and the \ci (2-1) transition is even closer to the CO (7-6) transition at 809 and 806 GHz, respectively.  Their proximity makes the line ratios \ci(1-0)/CO(4-3) and \ci(2-1)/CO(7-6) less affected by uncertainties in the baseline and corrections applied by the SECT.  We use RADEX models to produce line ratios over a range of kinetic temperatures and H$_2$ densities with a fixed column density for C of 5$\cdot$10$^{17}$ and 8$\cdot$10$^{18}$ for CO \citep{van2007}.  The temperatures vary from 10 to 100 K and the densities range from 10$^{1}$ to 10$^{8}$ cm$^{-3}$. We then use an error minimization routine to determine the conditions that best reproduce our observed line ratios. Uncertainties in the model solutions are based on 1 $\sigma$ variations in the line ratios.  If solutions showed no change, the grid spacing is reported as the uncertainty. The RADEX results can also be compared to temperatures derived directly from the ratio of the \ci(1-0) and (2-1) emission lines.  The results of the RADEX models can be seen in Table \ref{RADEXresults}, with the last column showing the temperatures derived solely from the \ci line ratio.

For the majority of galaxies in our sample, there is good agreement between the temperatures derived from the \ci line ratio and those inferred from the \ci/CO ratios from RADEX models.  However, one galaxy in our sample (NGC 2764) has line ratios which best match high temperature ($\sim$80 K), low density RADEX models, while the temperature derived from the \ci line ratio was moderate ($\sim$20 K). The discrepancy is largely motivated by the upper limit on the \ci(1-0) flux from NGC 2764, which in turn creates an upper limit on the \ci(1-0)/CO(4-3) ratio of 0.28.  Ratio values below this limit begin around 55 K, and at a low H$_2$ density RADEX can reproduce the observed \ci(2-1)/CO(7-6) ratio of 1.63.

Additionally, the PDRT models did not have solutions that could produce the observed values of \ci(2-1)/CO(1-0) for our ETGs.  Our ratios mostly fall between atomic hydrogen densities with log(n) of 4 or 5, but far below any G$_0$ contours as shown in Figure \ref{CICO_contour}.  However, RADEX models can reproduce our observed \ci(2-1)/CO(1-0) with a wide variety of conditions. Ratios for our ETGs vary between about 10-60.  At T=10 K, molecular hydrogen densities with log(n) between 4 and 8 and produce ratios of 10-11.  At temperatures of 20-60 K and beyond, it appears any ratio between 10 and 60 can be reproduced at some respective density. At T=20 the densities vary between log(n) of 2.3 to 8, above that they range from 2 to 3.5 or 4.  RADEX seems to be better suited for emission lines originating within molecular clouds, while the PDRT is more suited for the PDR layers on the surface of molecular clouds.

\begin{deluxetable*}{ lccccc }
\tablecaption{RADEX Modeling Results}
\tablehead{
 \colhead{Galaxy} & \colhead{$\frac{[CI](1-0)}{CO(4-3)}$} & \colhead{$\frac{[CI](2-1)}{CO(7-6)}$} & \colhead{T (K)} & \colhead{n (cm$^{-3}$)} & \colhead{T* (K)} } 
\tablecomments{Gas temperatures and H$_2$ densities obtained from comparing observed ratios to RADEX models. Uncertainties in the model solutions are based on 1 $\sigma$ variations in the line ratios.  If solutions showed no change, the grid spacing is reported as the uncertainty. *Temperature inferred from \ci line ratio in LTE.}
\startdata

NGC 1222 & $>$0.83 & 0.76 & 12.9$^{+1.4}_{-0.5}$ & 7.0$^{+0.1}_{-1.8}$ & 16.8$\pm$2.5\\
NGC 1266 & 0.52 & 0.66 & 30.8$^{+13.3}_{-6.9}$ & 3.7$^{+0.6}_{-0.6}$ & 25.8$\pm$3.6\\
NGC 2764 & $<$0.28 & 1.63 & 77.4$^{+17.6}_{-0.5}$ & 1.4$^{+0.1}_{-0.1}$ & $>$20.9\\
NGC 3665 & \nodata & $>$1.36 & \nodata & \nodata & $>$14.7\\
NGC 4459 & \nodata & $>$1.35 & \nodata & \nodata & $>$11.9\\
NGC 4526 & 0.72 & $>$1.83 & 19.5$^{+19.7}_{-9.0}$ & 1.6$^{+2.9}_{-0.4}$ & 15.7$\pm$2.0\\
NGC 4710 & $<$0.55 & $>$3.85 & \nodata & \nodata & $>$20.9\\
NGC 5866 & $<$1.32 & 2.23 & 54.1$^{+0.5}_{-40.5}$  & 2.1$^{+2.5}_{-0.1}$ & $>$13.7 \\

\enddata
\label{RADEXresults}
\end{deluxetable*}


\section{Conclusion}
\label{sec:conclusion}

We present measurements and analysis of line fluxes from \herschel SPIRE spectral observations of CO-rich early-type galaxies. These observations are some of the first ever detections of several FIR emission lines from ETGs including mid to high-J CO transitions as well as a pair of \ci transitions.  Our data reduction uses physically motivated corrections for the source size, thus we are not dependent of matching the SLW and SSW spectra in their overlapping region as there can be large random offsets (with respect to the continuum flux).  Additionally, we compare different but complementary methods for estimating the uncertainties in the fluxes of very weak lines in the SPIRE spectra.  The line fluxes are used to analyze the physical properties of the gas and compared to other galaxy types, mostly spirals and ULIRGs.

The mid J CO lines and the \ci emission lines in our sample of ETGs are relatively as bright (compared to the FIR flux) as those seen in other galaxies.  The CO SLEDs for ETGs mostly appear to be low excitation galaxies, peaking at an upper J of 4 or 5.  NGC 1266 is the exception, with a CO SLED that peaks at an upper J of 8. 

In previous work it was found that the \ci/FIR ratio exhibits a deficit similar to that observed for \cii/FIR (a deficit at high UV field strengths and hence higher FIR luminosities and warm FIR colors), while the CO lines are not affected by this deficit.  We agree that the CO/FIR ratio does not exhibit a deficit in ETGs, and we do not observe a deficit in the \ci/FIR either. The \ci(1-0)/FIR values for ETGs span a range of 10$^{-5}$ to 4.2$\cdot$10$^{-5}$, and the \ci(2-1)/FIR values span a range of 10$^{-5}$ to 3.8$\cdot$10$^{-5}$, roughly consistent with values found for other galactic morphologies.

Generally speaking, our sample of ETGs is largely consistent with spiral galaxies when considering the values of line ratios (i.e. \cii/FIR, \ci/FIR, \cii/\ci, etc.) and derived gas properties (i.e. temperatures, densities, and UV fields), suggesting that in ETGs with substantial amounts of molecular gas the behavior of the gas is quite similar to that of the gas in spiral galaxies.  Other recent observations of ETGs with radio jets and cooling flows have begun to find small quantities of molecular gas that has probably condensed out of thermal instabilities in the hot ionized gas (e.g. \citet{Russell, Temi2018}), and that molecular gas may behave differently from the kpc-scale disks we study here.

Using PDR models we find our ETGs are \textbf{mostly consistent with regular spiral galaxies.  When only the \cii, \oi, and FIR fluxes are used, five of the nineteen ETGs occupy a parameter space separate from the spiral galaxies and remaining ETGs. When \ci or CO(1-0) emission lines are included in the PDR models, all of our ETGs have solutions similar to spirals except for NGC4526, whose solution has a higher density than any other galaxy in the samples we studied.}

The use of \ci or CO(1-0) may be driving the PDR models to low G$_0$, high density solutions because neutral carbon and CO reside deeper inside the molecular clouds where densities are higher and fewer FUV photons are able to penetrate the gas. The inclusion of the \ci emission lines in particular seems to have the most dramatic effect on the model solutions.  The models predict roughly five times larger \ci(2-1)/CO(1-0) ratios at the values of n and G$_0$ inferred from the \cii, \oi, and FIR fluxes.

\textbf{Our PDR model results do not reproduce the scaling relation G$_0$ $\varpropto$ n$^{1.4}$ found in \citet{Mal01}, unless we do not apply a correction to the FIR flux.}

Though our sample of galaxies are all morphologically early-type, their inferred UV field strengths span two orders of magnitude from starbursting merger remnants to quiescent, relaxed galaxies.  In future work we plan to compare G$_0$ values found by PDR models to those derived from FUV observations.

Line ratios of the \ci transitions to neighboring CO transitions can be used to constrain the temperature and density of the emitting gas by comparing observed line ratios to those produced by RADEX models.  For the majority of our ETGs, temperatures found by RADEX models are similar to those derived from the \ci(1-0)/(2-1) line ratio.  This is roughly consistent with other samples of LIRGs and high redshift galaxies, which all have temperature derived from \ci in the range of 20-40 K, though the temperatures for ETGs tend to somewhat lower (several below 20 K, with some possibly as low as 12-15 K). However, the temperature from RADEX models for NGC 2764 ($\sim$ 77 K) is not in agreement with those found from \ci ratios, a possible indication that the conditions in the gas do not satisfy LTE.

For the most part the \ci line flux in our ETGs is proportional to CO(1-0), suggesting that either could be used for an estimate of the molecular mass.  Two of our galaxies show upper limits roughly ten times above the mass derived from CO, but they are still consistent with the result, and it is possible their spectra are too noisy.

The ETGs and nearby spirals confirm and extend one of the results from \citet{Lu}, which is that the \nii 205/CO(6-5) ratio shows a strong trend with the IRAS 60/100 \micron\ color.  In fact, the variation in this ratio among our ETGs is nearly a factor of 100.  Thus, they cannot both be equally good tracers of the star formation rate, at least not without some corrections.

\acknowledgments{{\it Acknowledgments:} }

This work is based on observations made with Herschel, a European Space Agency Cornerstone Mission with significant participation by NASA. Support for this work was provided by NASA through an award issued by JPL/Caltech.

Facilities: \facility{\herschel (PACS)}, \facility{\herschel (SPIRE)}


\clearpage
\begin{appendix}
\section{Uncertainties in the SPIRE Data}
\label{sec:montecarlo}

The high resolution SPIRE spectra that we study here, being obtained from a Fourier Transform technique, are effectively convolved with a sinc function of FWHM of 1.44 GHz.  At 500 GHz this FWHM corresponds to 860 km s$^{-1}$ and at 1400 GHz it corresponds to 306 km s$^{-1}$.  The processed data available from the archive are sampled at 0.3 GHz; thus, most of the galaxy lines we study are spectrally unresolved, but oversampled.  And, as discussed by \citet{Kam2016}, the oversampling has implications for estimates of the line flux uncertainties.  For very strong lines, of course, the line flux uncertainties may be dominated by the absolute calibration of the instrument, the baseline subtraction, corrections for the aperture size, and so on.  But for weak or marginally detected lines, an accurate treatment of the uncertainties requires some care.

There does not seem to be a clear consensus in the literature about how to treat weak lines in the SPIRE spectra.  For example, \citet{Rosenberg} adopt a straight 10\% flux uncertainty for the signal-to-noise component of the error budget, which is clearly too simplistic for weak lines.  \citet{Lu2017} report the formal flux uncertainties from the HIPE Spectrum Fitter tool, with additional quality flags derived from a comparison of the observed frequency and the known velocity.  \citet{Kam2016} carry out a Bayesian analysis, adding trial lines into the line-free residuals of the observed spectra to infer the distributions of line parameters that could reproduce the observations.  \citet{Rigopoulou} and \citet{Magdis} use a bootstrap technique that is appropriate for observations with many repeated scans, where subsets can be constructed and compared to each other.  In this paper we investigate the Bayesian method and compare to a simpler Monte Carlo method which can also account for the oversampling in the data.  We have not investigated the bootstrap method, though some of our data have enough repetitions that they would probably admit this kind of analysis.

For a simple Monte Carlo experiment to take account of the correlation in adjacent data points, we carry out an initial fit of the CO, \ci and \nii 205 lines in a spectrum.  We retain the fitted amplitudes, widths, and velocities of those lines in the appropriate sinc or sinc-Gauss models. We add random noise with similar statistical properties to the real data and re-fit.  After many realizations of the noise, the distribution of fitted amplitudes can be used to estimate the uncertainty in the fitted amplitude.  The noise is generated as a set of independent Gaussian random variables, convolved with the appropriate sinc function.  If the convolution kernel is normalized so that the sum of the square of its weights is 1, the convolution does not change the standard deviation of the noise, which can then be set to the desired level to match the line-free residuals in the data.

The results of this simple experiment show that the standard deviation of the fitted amplitudes is very close to the standard deviation in the line-free residuals of the original data.  For the sample size and numbers of iterations that we chose, the ratio of these quantities is 1.00 with a dispersion of 0.033 dex.  The factor of 1 comes from the fact that the spectral lines we are studying are mostly unresolved, so that there is effectively only one independent datapoint per line.  In the absence of the noise convolution step, so that the individual noise values are independent of each other, the standard deviation in the fitted amplitudes is almost exactly a factor of two lower.  In other words, the correlation between adjacent data points increases the uncertainties in fitted amplitudes by a factor of two; this value also makes sense, since the FWHM of the sinc convolution kernel is 4.77 times the sampling interval in the data.

The experiment thus suggests that the local rms in the spectrum is an appropriate estimate for the uncertainty in the line amplitude, at least in the cases of weak and unresolved lines.  Of course, the local rms in the spectrum can vary substantially with frequency; in our targets the variation is sometimes a factor of 3 between the low frequency end of SLW and the center of SLW.   For this experiment we capture the variation by measuring the local rms at several locations through the band and interpolating.  And we find that this variation is not reflected in the formal uncertainties reported by the built-in HIPE Spectrum Fitter routines, so we conclude that they are not adequately capturing uncertainties related to signal-to-noise issues for weak lines.  

For a consistency check, we also carried out a Bayesian uncertainty estimate for our targets using the method described by \citet{Kam2016}.  Rather than trying to simulate the noise, this method uses the observed line-free residuals in the data and places trial spectral lines at a various locations in the residuals to simulate the way the spectral line may interfere constructively or destructively with noise ripples.  For our targets, our estimates of the Bayesian uncertainties are only 18\% smaller than Kamenetzky's, with a dispersion of 0.12 dex (32\%).  As usual, the Bayesian method seems to be somewhat sensitive to what one assumes for the priors, such as the range of velocities, and how closely one sets the criterion for a match between the output ``measured'' flux and the original fitted flux.  Interestingly, this Bayesian method also reports that the uncertainty in the fitted amplitudes is equal to the local rms in the spectrum.  (The median ratio is 1.01, with a standard deviation of 0.12 dex or 32\%.)  The results remind us that the uncertainty estimates are themselves uncertain, of course, but at least the orthogonal Monte Carlo and Bayesian estimates are consistent with each other at this level.

\begin{turnpage}
\setcounter{table}{0}
\renewcommand{\thetable}{A\arabic{table}}
\begin{deluxetable*}{ lcccccccccccc }
\tablecaption{SECT Correction Factors}
\tablehead{
 \colhead{Galaxy} & \colhead{CO(4-3)} & \colhead{CO(5-4)} & \colhead{CO(6-5)} & \colhead{CO(7-6)} & \colhead{CO(8-7)} & \colhead{CO(9-8)} & \colhead{CO(10-9)} & \colhead{CO(11-10)} & \colhead{CO(12-11)} & \colhead{CO(13-12)} & \colhead{\ci(1-0)} & \colhead{\ci(2-1)} } 
\tablecomments{SECT correction factors for the CO and \ci transitions.  The uncorrected fluxes and uncertainties can be obtained by dividing by the factors given in this table.}
\startdata

NGC 1222&1.02&1.03&1.04&1.02&1&1.18&1.23&1.23&1.25&1.26&1.03&1.02\\
NGC 1266&1&1&1.01&1&1&1.02&1.02&1.02&1.02&1.02&1&1\\
NGC 2764&1.01&1.01&1.01&1.01&1&1.04&1.05&1.06&1.06&1.06&1.01&1.01\\
NGC 3665&1.01&1.02&1.03&1.02&1&1.12&1.14&1.15&1.15&1.16&1.02&1.02\\
NGC 4459&1.02&1.03&1.04&1.02&0.99&1.17&1.2&1.21&1.22&1.23&1.03&1.02\\
NGC 4526&1.02&1.04&1.05&1.03&0.99&1.22&1.27&1.28&1.29&1.3&1.03&1.03\\
NGC 4710&1.02&1.03&1.04&1.02&0.99&1.18&1.21&1.22&1.23&1.24&1.03&1.02\\
NGC 5866&1.01&1.02&1.02&1.01&1&1.09&1.11&1.12&1.12&1.13&1.02&1.01\\
NGC 7465&1.02&1.03&1.05&1.02&0.99&1.2&1.24&1.25&1.26&1.28&1.03&1.02\\

\enddata
\label{CorrFactors}
\end{deluxetable*}
\end{turnpage}
\setcounter{table}{1}
\renewcommand{\thetable}{A\arabic{table}}

\begin{deluxetable*}{ lcccccc }
\tablewidth{0pt}
\tablecaption{PDR Toolbox Results}
\tablehead{
 \colhead{} & \multicolumn{2}{c}{w/ \ci and/or CO} & \multicolumn{2}{c}{Only \cii, \oi, and FIR} & \multicolumn{2}{c}{w/ Uncorrected FIR} \\ \colhead{Galaxy} & \colhead{n} & \colhead{G$_0$} & \colhead{n} & \colhead{G$_0$}& \colhead{n} & \colhead{G$_0$}  } 
\tablecomments{ Results from the PDR Toolbox.  Columns 2 and 3 reflects solutions including CO(1-0) and the \ci lines where available.  Columns 4 and 5 are solutions using only \cii, \oi, and FIR fluxes.  Columns 6 and 7 are solutions using only \cii, \oi, and FIR, but without correcting the FIR flux.  The first nine rows are galaxies with \ci observations.} 
\startdata

NGC 1222&4.25&2.75&4.25&1.5&3.5&2.5\\
NGC 1266&4.25&1.75&3&3&2.75&3.25\\
NGC 2764&3.5&2.25&3.75&1.75&3.25&2.25\\
NGC 3665&4.25&2.25&3.75&2.75&3.5&3.25\\
NGC 4459&4.5&1.25&3.25&3.5&3&3.75\\
NGC 4526&5.25&0.75&4.25&4.75&4&5\\
NGC 4710&3&2.5&3&2.5&3&3\\
NGC 5866&4&1.5&1.75&2.5&1.5&2.75\\
NGC 7465&3.5&2.5&4&2&3.25&2.5\\
IC 0676&4.25&2&3.75&2.25&3.25&2.75\\
IC 1024&3.5&2&3.75&2&3.25&2.5\\
NGC 3032&4&1.25&3&2.25&2.75&2.75\\
NGC 3489&3.75&1.5&2.75&2.5&1&1.75\\
NGC 3607&\nodata & \nodata &1.75&2.25&4.25&-0.25\\
NGC 3626&3.5&1.75&3.5&2.5&3&2.75\\
NGC 4150&4.25&1.25&3&3&2.75&3.25\\
NGC 4429&4&1&1.25&1.75&1.75&2.5\\
NGC 4435&4&1.5&1.75&2.25&1.5&2.5\\
NGC 4694&3.25&1.5&3.25&2.25&2.75&2.5\\
  IC 2574 &\nodata&\nodata&5&1.75&3.75&2.75\\
 NGC 0337 &\nodata&\nodata&3.5&2&3&2\\
 NGC 0628 &\nodata&\nodata&3.5&2&2.75&1.75\\
 NGC 0855 &\nodata&\nodata&3.75&2&3&2\\
 NGC 0925 &\nodata&\nodata&3&1.5&2.25&1.25\\
 NGC 1291 &\nodata&\nodata&2.75&2.5&1.75&1.75\\
 NGC 1316 &\nodata&\nodata&3&2.5&2.5&2.25\\
 NGC 1482 &\nodata&\nodata&3.5&2.5&4.25&0.25\\
 NGC 1512 &\nodata&\nodata&3&2.25&2.25&1.75\\
 NGC 2798 &\nodata&\nodata&3.5&2.75&3&2.5\\
 NGC 2915 &\nodata&\nodata&4.25&1&3.75&1.75\\
 NGC 2976 &\nodata&\nodata&3.5&2&3&2\\
 NGC 3049 &\nodata&\nodata&3.75&2&3&2\\
 NGC 3077 &\nodata&\nodata&3.25&2.25&4&0.25\\
 NGC 3184 &\nodata&\nodata&3&2.25&1.25&1\\
 NGC 3190 &\nodata&\nodata&3&3&2.25&2.5\\
 NGC 3198 &\nodata&\nodata&2.75&2.25&2&1.75\\
 NGC 3265 &\nodata&\nodata&4&2&3.25&2\\
 NGC 3351 &\nodata&\nodata&3&2.5&2.5&2.25\\
 NGC 3621 &\nodata&\nodata&3&2&2.5&1.75\\
 NGC 3627 &\nodata&\nodata&3&2.25&4&0.25\\
 NGC 3773 &\nodata&\nodata&3.5&2&3&1.75\\
 NGC 3938 &\nodata&\nodata&3&2.25&4&0.25\\
 NGC 4254 &\nodata&\nodata&2.75&2&2.25&1.75\\
 NGC 4321 &\nodata&\nodata&4.25&0.25&2.25&2\\
 NGC 4536 &\nodata&\nodata&3.75&2.25&3&2.25\\
 NGC 4559 &\nodata&\nodata&3.25&2&2.5&1.75\\
 NGC 4569 &\nodata&\nodata&3.25&2.5&2.75&2.25\\
 NGC 4579 &\nodata&\nodata&3.5&2.25&3&2.25\\
 NGC 4625 &\nodata&\nodata&4&0.5&2.75&1.75\\
 NGC 4631 &\nodata&\nodata&3&2&4&0.25\\
 NGC 4736 &\nodata&\nodata&3.25&2.25&2.5&2\\
 NGC 5408 &\nodata&\nodata&4.5&1.25&3.75&2\\
 NGC 7331 &\nodata&\nodata&2.75&2.25&1.5&1.25\\
 NGC 7793 &\nodata&\nodata&3.25&2&2.75&1.75\\
  IC 0342 &\nodata&\nodata&3.25&2.75&2.75&2.5\\
 NGC 1097 &\nodata&\nodata&4.5&0.25&2.75&2.75\\
 NGC 2146 &\nodata&\nodata&3.5&2&3&2\\
 NGC 3521 &\nodata&\nodata&3&2.5&2.25&2\\
 NGC 4826 &\nodata&\nodata&3&2.75&1.5&1.75\\
 NGC 5055 &\nodata&\nodata&3.25&3.5&2.75&3.25\\
 NGC 5713 &\nodata&\nodata&3.5&2.5&4.25&0.25\\
 NGC 6946 &\nodata&\nodata&3&2.5&2.5&2.25
\enddata
\label{PDRT_table}
\end{deluxetable*}

\end{appendix}
%

\FloatBarrier
\newpage
\bibliographystyle{aj}


\end{document}